\documentclass[11pt,fleqn]{article}       
\usepackage{geometry}
\usepackage[T1]{fontenc}
\usepackage[utf8]{inputenc}
\usepackage{authblk}
\usepackage{mathrsfs}
\usepackage[font=small,labelfont=bf]{caption}
\usepackage{graphicx}
\usepackage{multirow}
\usepackage{amsmath,amssymb,amsfonts}
\usepackage{verbatim}
\usepackage{bm}
\usepackage{color}
\usepackage{ulem}
\usepackage{mathtools}
\usepackage{cite}
\usepackage{colortbl}
\usepackage{cancel}
\definecolor{light-gray}{gray}{0.85}
\usepackage[pdftex,colorlinks,pdfpagelabels]{hyperref}
\hypersetup{
   bookmarksnumbered,
   citecolor={blue},
   linkcolor={blue},
   urlcolor={blue},
   filecolor={blue}
} 

\newcommand{\eVdist}{\kern-0.06em}

\newcommand{\kev}{\:\text{ke\eVdist V}}

\newcommand{\gev}{\:\text{Ge\eVdist V}}

\newcommand{\AddrTexas}{%
\textit{Department of Physics, The University of Texas at Austin, Austin, 78712 TX, USA}
}
\newcommand{\AddrStockholm}{
\textit{Oskar Klein Center for Cosmoparticle Physics, University of Stockholm, 10691 Stockholm, Sweden}
}
\newcommand{\AddrNordita}{
\textit{Nordita, KTH Royal Institute of Technology and Stockholm University, 10691 Stockholm, Sweden}
}

\usepackage{fancyhdr}
 \usepackage[bottom]{footmisc}
\fancypagestyle{plain}{%
    \fancyhead[R]{UTWI-06-2023, NORDITA-2023-003}
    
}
\date{}
\title{Dark Matter and Gravity Waves from a Dark Big Bang
}
\author[1,2,3]{Katherine Freese\thanks{ktfreese@utexas.edu}}
\author[1,2]{Martin Wolfgang Winkler\thanks{martin.winkler@austin.utexas.edu}}
\affil[1]{\AddrTexas}
\affil[2]{\AddrStockholm}
\affil[3]{\AddrNordita}

\begin{document}
\maketitle
\vspace*{0mm}
\begin{abstract}
The Hot Big Bang is often considered as the origin of all matter and radiation in the Universe. Primordial nucleosynthesis (BBN) provides strong evidence that the early Universe contained a hot plasma of photons and baryons with a temperature $T>\text{MeV}$. However, the earliest probes of dark matter originate from much later times around the epoch of structure formation. In this work we describe a scenario in which dark matter (and possibly dark radiation) can be formed around or even after BBN in a second Big Bang which we dub the ``Dark Big Bang''. The latter occurs through a  phase transition in the dark sector which transforms dark vacuum energy into a hot dark plasma of particles; in this paper we focus on a first-order phase transition for the Dark Big Bang. The correct dark matter abundance can be set by dark matter cannibalism or by pair-annihilation within the dark sector followed by a thermal freeze-out. Alternatively ultra-heavy ``dark-zilla'' dark matter can originate directly from bubble collisions during the Dark Big Bang. We will show that the Dark Big Bang is consistent with constraints from structure formation and the Cosmic Microwave Background (CMB) if it occurred when the Universe was less than one month old, corresponding to a temperature in the visible sector above $\mathcal{O}$(keV). While the dark matter evades direct and indirect detection, the Dark Big Bang gives rise to striking gravity wave signatures to be tested at pulsar timing array experiments. Furthermore, the Dark Big Bang allows for realizations of self-interacting and/or warm dark matter which suggest exciting discovery potential in future small-scale structure observations.
\end{abstract}
\clearpage

\section{Introduction}

According to the cosmological standard model, the very early Universe went through an epoch of inflation~\cite{Guth:1980zm} -- a rapid expansion of space driven by vacuum energy. The origins of matter and radiation lie in the Hot Big Bang which terminates inflation and releases the vacuum energy into a hot plasma of particles. The latter contains the photons, leptons and quarks of our visible Universe, and, in the standard picture, also the dark matter -- for instance in the form of Weakly Interacting Massive Particles (WIMPs).
However, there is no genuine reason for a common origin of visible and dark matter beyond simplicity. In fact, while the presence of photons and baryons at very early stages is well-established by the successful theory of BBN (see e.g.~\cite{Schramm:1997vs,Steigman:2007xt,Cyburt:2015mya}), there exist no probes of dark matter prior to the time when observable scales re-entered the horizon (at keV temperatures of the Universe) and the impact of dark matter on the first structures started to manifest. Furthermore -- despite excessive experimental searches over decades -- no direct non-gravitational interactions between visible and dark matter have been detected (see e.g.~\cite{Fermi-LAT:2016uux,XENON:2018voc}).

\begin{figure}[h!]
\begin{center}
\includegraphics[width=0.81\textwidth]{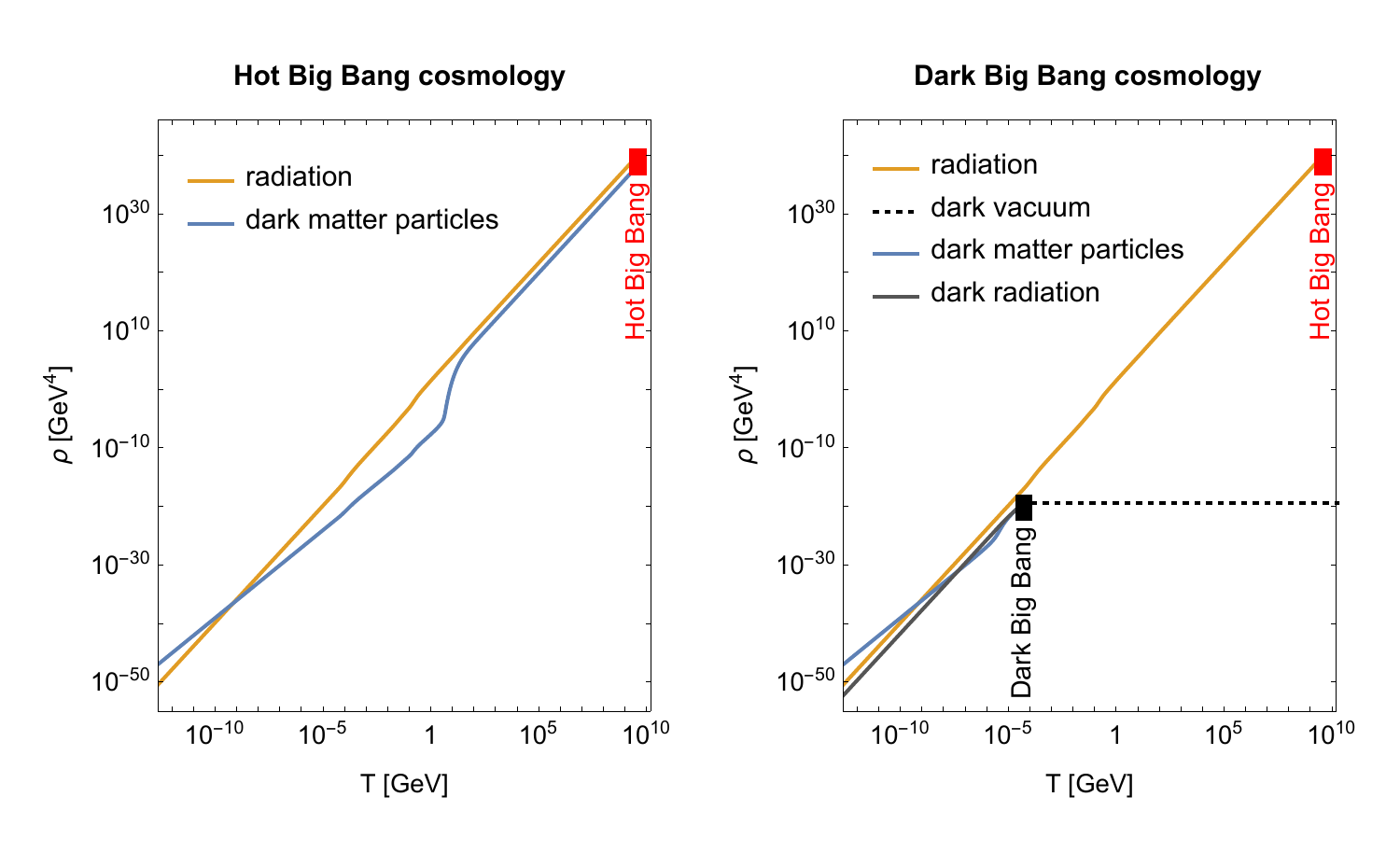}
\end{center}
\vspace{-10mm}
\caption{Typical temperature evolution of energy densities in a standard Hot Big Bang cosmology (left panel) and in the Dark Big Bang scenario studied in this work (right panel). The curves for dark matter indicate the energy density carried by dark matter particles (in the left panel, an illustrative curve where dark matter particles are initially relativistic with energy density $\propto T^4$ and finally non-relativistic,  $\propto T^3$).
In Hot Big Bang cosmology all matter and radiation is produced in the Hot Big Bang. In the Dark Big Bang scenario the Hot Big Bang only produces the visible matter and radiation, while the Dark Big Bang induces the dark matter and (possibly) dark radiation.}
\label{fig:hotdark}
\end{figure}

In this light, we will present an alternative cosmological scenario in which the visible and the dark (matter) sector are completely decoupled (other than through gravity).\footnote{Variants of the our Dark Big Bang scenario with small couplings between the two sectors would be interesting to investigate in the future.}
The Hot Big Bang only induces visible radiation and matter, but no dark matter at all. However, while the dark sector is cold at first, it contains a small amount of vacuum energy which is initially highly subdominant compared to the radiation density of the Universe. 
Because vacuum energy does not redshift, the dark vacuum contribution can  become significant at a later stage of the Universe, although it never dominates the energy density of the Universe. When the dark vacuum finally decays in a dark phase transition, it can induce significant amounts of dark matter and possibly dark radiation. Due to the analogy with the Hot Big Bang -- which transfers the inflationary vacuum energy into visible particles -- we call this process the Dark Big Bang~\cite{Freese:2022qrl}. 
In previous work we  pointed out the idea of the Dark Big Bang in~\cite{Freese:2022qrl}, and in this paper we explore the scenario in more detail.
The Dark Big Bang scenario is compared to a standard Hot Big Bang cosmology in Fig.~\ref{fig:hotdark}.

We have been particularly interested in answering the following question: what is the latest time at which the Dark Big Bang could take place in the history of the Universe? Clearly, for purely gravitational couplings between the dark and visible sectors, the Dark Big Bang can occur after BBN without spoiling the light element abundances. But another key issue is that the dark matter must pick up the right adiabatic perturbations required for structure formation. Indeed, we will show that the leading constraints on the time of the Dark Big Bang arise from structure formation and allow for a Dark Big Bang as late as $\mathcal{O}(\text{month})$ after the Hot Big Bang (corresponding to a redshift of $z\simeq 3\times 10^6$). We note, however, that the Dark Big Bang cannot be pushed to an epoch as late as matter-radiation equality at $z=3500$ (as preferred by early dark energy solutions to the Hubble tension~\cite{Karwal:2016vyq,Poulin:2018cxd}) without spoiling Lyman-$\alpha$ and CMB observations.\footnote{We note, however, that it is possible to generate a fraction of the dark matter density shortly before matter-radiation equality~\cite{Niedermann:2021ijp,Niedermann:2021vgd}.}

While many of our findings will be applicable to any type of Dark Big Bang phase transition, we will focus on the case, where the Dark Big Bang is associated with a first-order phase transition.
Several dark matter realizations connected to a first-order phase transition have previously been discussed in the literature which include the formation of heavy dark matter by bubble collisions~\cite{Watkins:1991zt,Chung:1998ua,Kolb:1998ki,Falkowski:2012fb,An:2022toi} or bubble expansion~\cite{Azatov:2021ifm}, asymmetric dark matter~\cite{Shelton:2010ta,Petraki:2011mv,Baldes:2017rcu,Hall:2019rld}, Q-ball dark matter~\cite{Krylov:2013qe,Huang:2017kzu}, Fermi ball dark matter~\cite{Hong:2020est,Kawana:2021tde,Kawana:2022lba}, quark nugget dark matter~\cite{Witten:1984rs,Frieman:1990nh,Zhitnitsky:2002qa,Oaknin:2003uv,Lawson:2012zu,Atreya:2014sca,Bai:2018vik,Bai:2018dxf}, filtered dark matter~\cite{Baker:2019ndr,Chway:2019kft} and primordial black hole dark matter~\cite{Hawking:1982ga,Crawford:1982yz,Kodama:1982sf,Moss:1994pi,Freivogel:2007fx,Johnson:2011wt,Kusenko:2020pcg,Baker:2021nyl,Jung:2021mku}\footnote{In some of the listed references the primordial black holes only account for a fraction of the dark matter.}. Our Dark Big Bang proposal differs from these complementary ideas because we are considering the false vacuum decay into a dark particle plasma within a decoupled previously cold dark sector (such that the Dark Big Bang is the dark sector analogue of the Hot Big Bang).

Specifically, we will introduce a dark sector scalar field which initially populates a metastable minimum in its potential. Later, during the radiation-dominated epoch of the Universe, the dark scalar tunnels into the true minimum~\cite{Coleman:1977py} and initiates the Dark Big Bang: bubbles of true vacuum form at random nucleation sites, expand, collide and produce dark matter and dark radiation. 

We will investigate the dark matter production by the bubble collisions themselves~\cite{Watkins:1991zt,Falkowski:2012fb} and by thermal processes in the dark plasma emerging from the Dark Big Bang. Both production modes will be shown to be capable of inducing the correct dark matter relic density observed in our Universe today. Viable dark matter from the Dark Big Bang can span an enormous mass range of $\sim 20$ orders of magnitude: while the bubble collisions can successfully generate ultra-heavy dark-zilla dark matter with a mass as large as $m_\chi\sim 10^{12}\:\text{GeV}$, thermal scattering and freeze-out in the dark plasma can provide realizations of dark WIMP dark matter or dark cannibal dark matter as light as $m_\chi\sim\text{keV}$.

The dark matter from the Dark Big Bang evades direct and indirect detection experiments because it only gravitationally couples to ordinary matter. However, we will show that realizations of warm dark matter, or self-interacting dark matter naturally arise. Furthermore, the dark radiation produced in the Dark Big Bang can increase the effective neutrino number $N_{\text{eff}}$ in the early Universe. Hence, exciting signatures of the Dark Big Bang can arise in the small-scale structure of the Universe (see e.g.~\cite{Spergel:1999mh,Bode:2000gq,Tulin:2017ara,Bullock:2017xww}).

Besides the dark matter signatures, the Dark Big Bang induces significant amounts of gravitational radiation by the collision of true-vacuum bubbles during the phase transition~\cite{Witten:1984rs,Hogan:1986qda,Kosowsky:1992rz,Kosowsky:1992vn}. The particularly intriguing case of a Dark Big Bang around or after BBN would imply a peak frequency of the gravitational wave spectrum in the nHZ or sub-nHz-regime. We will show that ongoing (North American~\cite{NANOGrav:2020bcs}, European~\cite{Chen:2021rqp}, Parkes~\cite{Goncharov:2021oub} and International Pulsar Timing Arrays~\cite{Antoniadis:2022pcn}) and upcoming (Square Kilometre Array~\cite{Dewdney:2009}) gravitational wave searches by pulsar timing arrays exhibit a striking potential to discover the Dark Big Bang phase transition. 

We leave to future work variations of the discussed scenario: generalization to phase transitions that are not first-order as well as the possibility of including small couplings between the dark and the visible sector which could give rise to interesting signatures in a variety of data sets. 

This work is organized as follows: in Sec.~\ref{sec:scenario} we describe the cosmological evolution and the initial conditions of the Dark Big Bang scenario. Furthermore, we use observational data to constrain the strength of the Dark Big Bang. In Sec.~\ref{sec:structure} we investigate the dark matter perturbations. We will show that -- if the Dark Big Bang occurs when the Universe is less than a month old  - the dark matter receives the desired adiabatic perturbations to support structure formation, while unwanted isocurvature and peaked perturbations are absent (or occur only on unobservably small scales). Then, in Sec.~\ref{sec:model} we introduce an explicit Dark Big Bang model realization and study the false vacuum decay rate which determines the time of the Dark Big Bang. In Sec.~\ref{sec:dmproduction} we investigate the dark matter production by the Dark Big Bang and discuss a number of scenarios which can reproduce the observed dark matter density. In Sec.~\ref{sec:gravitywaves}, we calculate the gravitational wave signal of the Dark Big Bang and show the discovery potential of gravitational wave searches with pulsar timing arrays. Finally, Sec.~\ref{sec:conclusion} contains our summary and conclusions.

\section{A Dark Big Bang in the Early Universe}\label{sec:scenario}

\begin{figure}[htp]
\begin{center}
\includegraphics[width=0.5\textwidth]{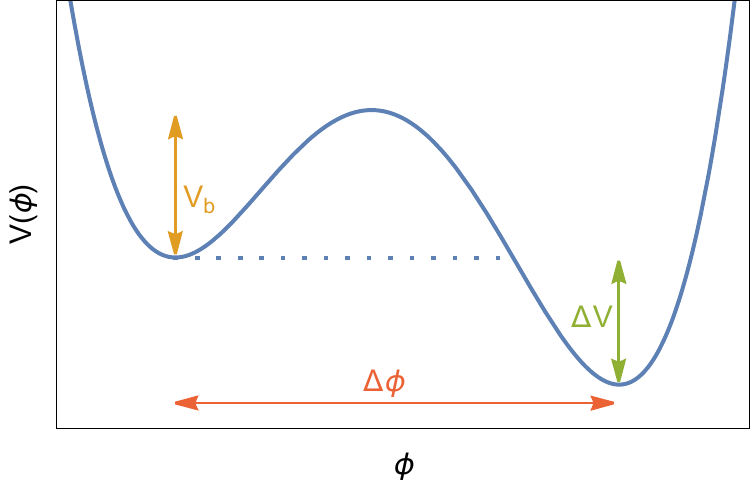}
\end{center}
\caption{Potential of the scalar field $\phi$ featuring two non-generate minima separated by a barrier of height $V_b$. The potential and field difference between minima are denoted by $\Delta V$ and $\Delta\phi$ respectively. Initially, $\phi$ is trapped in the higher-energy minimum. It later tunnels into the lower minimum, thereby producing the dark matter of the Universe.}
\label{fig:potential}
\end{figure}

We consider a dark sector containing a real scalar field $\phi$ and a stable particle $\chi$. In the following $\phi$ will be the field which triggers the Dark Big Bang phase transition, while $\chi$ will play the role of the dark matter. Optionally, additional massless (or very light) degrees of freedom $\xi_i$ coupling to $\phi,\,\chi$ may be present in the dark sector playing the role of dark radiation. However, we assume that all dark sector fields are decoupled from ordinary matter other than through gravity. The potential $V(\phi)$ of the scalar $\phi$ is chosen to feature two non-degenerate minima. We denote the potential difference, field distance and barrier height between the minima by $\Delta V$, $\Delta\phi$ and $V_b$ respectively (see Fig~\ref{fig:potential}).

\subsection{Cosmological Evolution of the Dark Sector}\label{sec:darkcosmology}
Turning to the early Universe, we consider a standard cosmology with inflation ending in a Hot Big Bang (=reheating) which creates a thermal plasma of Standard Model (SM) particles. The reheating temperature $T_R$ depends on the details of the underlying inflation model, but a lower bound $T_R > 2\: \text{MeV}$ is set by BBN~\cite{Hannestad:2004px,Hasegawa:2019jsa}. We assume that reheating to dark sector particles is suppressed such that the energy density of the Universe $\rho_{\text{tot}}$ is (strongly) dominated by the SM radiation bath. The dark sector does never reach thermal equilibrium with the visible sector throughout the cosmological evolution.  We note that, throughout the paper, we will use the terminology ``radiation domination'' to mean domination by the visible sector radiation plasma (as opposed to the dark radiation in the dark sector).

At the beginning of the radiation-dominated epoch after inflation we thus have
\begin{equation}
\rho_{\text{tot}}\simeq \rho_r=\frac{\pi^2}{30} g_{\text{eff}}(T)\, T^4\,,
\end{equation}
where $\rho_r$ denotes the radiation energy density (of the visible sector) and $T$, $g_{\text{eff}}$ the temperature, number of relativistic degrees of freedom in the visible sector.

However, we take $\phi$ to populate the higher metastable minimum after inflation throughout the entire (observable) Universe -- an assumption we will motivate in Sec.~\ref{sec:initial}. The energy density of the dark sector $\rho_{\text{DS}}$ is initially given by the vacuum energy of $\phi$,
\begin{equation}
\rho_{\text{DS}}\simeq \rho_\phi = \Delta V\,.
\end{equation}
We neglected the initial dark radiation contribution to $\rho_{\text{DS}}$ following our earlier assumption that the Hot Big Bang only (or dominantly) heats the visible sector. While initially also $\rho_\phi\ll \rho_r $ (otherwise $\phi$ would be the inflaton), the ratio increases because radiation energy redshifts as $a^{-4}$ with the scale factor $a$, whereas $\rho_\phi$ remains constant as long as $\phi$ populates the false vacuum. Consequently, the dark sector vacuum energy can become significant at a later stage of the Universe. It should, however, never dominate the energy content in order not to spoil the cosmological evolution (see Sec.~\ref{sec:strength}).\footnote{If the false vacuum dominates the energy density prior to its decay, the Universe would typically still be dominated by the dark sector energy density around the epoch of last scattering which is excluded by CMB constraints. An exception is the case where the false vacuum energy density is converted into a cosmic fluid which redshifts faster than radiation. However, this case is strongly constrained by requiring the observed adiabatic fluctuations of radiation and matter and will not be considered in this work.}

Later $\phi$ tunnels into its lower minimum triggering a first-order phase transition which we dub the ``Dark Big Bang''.\footnote{One could also imagine the case of many Dark Big Bangs if the dark sector undergoes a series of first-order phase transitions instead of just one -- related to the ideas of chain inflation~\cite{Freese:2004vs,Freese:2005kt,Ashoorioon:2008pj,Winkler:2020ape,Freese:2021noj} and chain early dark energy~\cite{Freese:2021rjq}.} Bubbles of true vacuum form at random nucleation sites and expand into the sea of false vacuum. 

Since the dark sector is cold prior to the Dark Big Bang, and since $\phi$ does not couple to the visible sector, the potential $V(\phi)$ is not affected by thermal corrections prior to the phase transition. This distinguishes the Dark Big Bang scenario from complementary approaches in the literature which consider a first-order phase transition in a thermalized dark sector (see e.g.~\cite{Jaeckel:2016jlh,Addazi:2016fbj,Breitbach:2018ddu,Wang:2022akn}).\footnote{The case of quantum tunneling in a cold dark sector is briefly discussed in~\cite{Ellis:2018mja,Fairbairn:2019xog}.} We can, hence, apply the formalism for quantum tunneling in vacuum. The bubble nucleation rate per volume is determined by~\cite{Coleman:1977py}
\begin{equation}\label{eq:bounce1}
\Gamma = A e^{-S}\,,
\end{equation}
where $S$ denotes the Euclidean action of the bounce solution interpolating between the two minima in the potential. The prefactor $A$ incorporates quantum fluctuations about the classical action~\cite{Callan:1977pt}. It can be approximated by
\begin{equation}\label{eq:bounce2}
A\simeq m^4 \left(\frac{S}{2\pi}\right)^2\,,
\end{equation}
where $m$ stands for the mass of the tunneling scalar field evaluated in the false vacuum. Due to the absence of temperature effects on the tunneling rate $\Gamma$ is a time-independent quantity.  

Because of the exponential suppression of the tunneling rate, the false vacuum can easily be very long-lived. The mean lifetime of the Universe in the false vacuum $t_*$ is determined by $\Gamma \mathcal{V}_4(t_*)=1$~\cite{Guth:1979bh,Guth:1981uk}. Here $\mathcal{V}_4(t)$ stands for the spacetime volume of the past lightcone at the time $t$ (at a random position in space). The product $\Gamma \mathcal{V}_4$ thus measures the mean number of bubble nucleation sites in the past lightcone. Assuming that the energy density of $\phi$ is still subdominant to the SM radiation bath at the phase transition, we can (approximately) employ the expansion history of a radiation-dominated Universe to find~\cite{Freese:2022qrl}
\begin{equation}\label{eq:tstar}
t_*\simeq\left(\frac{105}{8\pi\,\Gamma}\right)^{1/4}\simeq 1.4 \times \Gamma^{-1/4}\, \,\,\,\,\,\,\,\, {\rm (time \,\, of \,\, the \,\, phase \,\, transition)}.
\end{equation}
The temperature of the SM plasma at the time $t_*$ of the phase transition is given by\footnote{Notice that the temperature of the dark sector is generically different from $T_*$ since both sectors are never in equilibrium. The temperature of the SM plasma only affects the dark sector by setting the expansion rate of the Universe.}
\begin{equation}\label{eq:Tstar}
T_* \simeq \left(\frac{45}{2\pi^2 }\right)^{1/4}\left(\frac{M_P^2}{g_{\text{eff}}(T_*)\,t_*^2}\right)^{1/4}
\simeq 0.2\kev \left(\frac{3.4}{g_{\text{eff}}(T_*)}\right)^{1/4} \left(\frac{\text{yr}}{t_*}\right)^{1/2}\,. 
\end{equation}
Throughout the paper, we will take the subscript $*$ to indicate that a quantity is evaluated at the time of the Dark Big Bang phase transition.

We can also estimate the duration of the phase transition $\beta^{-1}$ by considering how fast the number of bubble nucleation sites in the past lightcone increases with time (evaluated at the mean transition time)~\cite{Freese:2022qrl},
\begin{equation}\label{eq:beta}
\beta = \left.\frac{1}{\Gamma \mathcal{V}_4}\frac{d(\Gamma \mathcal{V}_4)}{dt}\right|_{t=t_*} = \left.\frac{\dot{\mathcal{V}}_4}{\mathcal{V}_4}\right|_{t=t_*}=\frac{4}{t_*}=8H_*\,,
\end{equation}
where we employed that the tunneling rate is time-independent and introduced the Hubble rate at the Dark Big Bang $H_*$. We comment that in most of the literature on first-order phase transitions one finds the definition $\beta=\dot{\Gamma}/\Gamma$. This definition, however, implicitly relies on the assumption that the tunneling field $\phi$ couples to the surrounding plasma. In the presence of such couplings the tunneling rate becomes time-dependent and one usually has $\dot{\Gamma}/ \Gamma \gg \dot{\mathcal{V}}_4/\mathcal{V}_4$ such that Eq.~\eqref{eq:beta} would indeed yield $\beta\simeq \dot{\Gamma}/\Gamma$. If $\phi$ is decoupled from the thermal plasma -- as in the scenario we described -- the tunneling rate is time-independent and the duration of the phase transition is controlled by the Hubble expansion rate.

Similar as the Hot Big Bang creates a hot plasma of visible sector particles, the Dark Big Bang heats up the dark sector. During the phase transition, the expanding bubble walls carry the energy previously contained in the false vacuum. Once the bubbles collide, the energy is released into the formation of dark sector particles and gravity waves
\begin{equation}
\rho_{\text{DS}}=
\begin{cases}
\rho_\phi & \text{for } t< t_*\,, \\ \rho_{\text{DR}} + \rho_\chi + \rho_{\text{GW}}  & \text{for } t\geq t_*\,,
\end{cases}
\end{equation}
with $\rho_\phi=\Delta V$, and we require continuity at $t_*$. Here we assumed that the entire false vacuum energy is released in the phase transition, i.e.\ we neglected any backreaction of the produced dark sector particles on the potential of the tunneling field.\footnote{The potential $V(\phi)$ typically exhibits a dependence on the temperature $T_{\text{DS}}$ of the dark sector particles generated in the phase transition. Therefore, a non-vanishing energy density may remain in the (thermal) potential after the phase transition. This contribution is, however, typically negligible in our Dark Big Bang scenario unless the dark sector particles couple strongly to the tunneling field.} Furthermore, we took the phase transition as instantaneous thus neglecting the time interval in which the energy is (partly) stored in the bubble walls. The error we introduce by this approximation (which we will only use in the context of energy densities) is relatively small since the phase transition only lasts a fraction of a Hubble time as shown in Eq.~\eqref{eq:beta}. After the transition, the dark sector energy density is shared among dark radiation (DR), dark matter particles ($\chi$) and gravity waves\footnote{For convenience, we count gravity waves as part of the dark sector energy density such that the total dark sector energy density is preserved at the phase transition.} (GW):
\begin{enumerate}
\item The dark radiation plasma is comprised of light dark sector degrees of freedom (if present). These typically reach thermal equilibrium among themselves (but not with the SM bath) quickly after the Dark Big Bang~\cite{Watkins:1991zt}. The dark radiation density redshifts as $\rho_{\text{DS}}\propto a^{-4}$ with the scale factor of the Universe.

\item Dark matter particles $\chi$ are generated non-thermally by the bubble collisions~\cite{Watkins:1991zt,Falkowski:2012fb} and/or thermally by scattering processes in the dark radiation plasma. While thermal production ceases for $m_\chi\gg T_{\text{DS}}$, heavier dark matter particles can still efficiently be generated by the colliding bubbles due to their Lorentz boost. In Sec.~\ref{sec:dmproduction}  we will discuss a variety of dark matter candidates in more detail.  After dark matter number changing processes are frozen out, $\rho_{\chi}\propto a^{-3}$ ($\rho_{\chi}\propto a^{-4}$) in the non-relativistic (highly relativistic) regime.

\item The energy density transferred into gravity waves at the phase transition can be estimated as (see e.g.~\cite{Caprini:2007xq})
\begin{equation}\label{eq:rhogwstar}
\rho_{\text{GW},*} \sim  3\left(\frac{H_*}{\beta}\right)^2 \frac{\rho_\phi^2}{\rho_{r}(T_*)}\simeq 0.05\frac{\rho_\phi}{\rho_{r}(T_*)}\rho_\phi\,.
\end{equation}
Since the Universe should not be vacuum-dominated at the transition, i.e. $\frac{\rho_\phi}{\rho_{r}(T_*)} < 1$, gravity waves make up less than $5\%$ of the dark sector energy density after the transition. The gravity wave energy density redshifts as $\rho_{\text{GW}}\propto a^{-4}$.
\end{enumerate}
We have implicitly assumed that there is always some light degree of freedom available in the dark sector to which the bubble walls can efficiently decay. If this is not the case, the colliding bubble condensate can potentially be long-lived and a description in terms of an effective fluid dominated by kinetic energy and small-scale anisotropic stress may arise~\cite{Niedermann:2020dwg}. We exclude such a scenario of inefficient particle production since it appears to be inconsistent with the generation of all dark matter in the phase transition (which is our definition of the Dark Big Bang).

In the presence of light dark sector degrees of freedom which reach a thermal equilibrium state quickly after the Dark Big Bang we can assign a dark sector temperature $T_{\text{DS}}$ to the resulting dark plasma. Furthermore, we can define the ``dark reheating temperature'' $T_{\text{DS},*}$ as the dark plasma temperature right after the Dark Big Bang. Since gravitational waves are subdominant (see above), we can approximate,
\begin{equation}\label{eq:darkreheating}
\rho_\phi\simeq \frac{\pi^2}{30} g_{\text{DS}}(T_{\text{DS,*}})\, T_{\text{DS,*}}^4\,,
\end{equation}
where $g_{\text{DS}}$ counts the number of light dark sector degrees of freedom. If $m_\chi <T_{\text{DS,*}}$ the dark matter particle $\chi$ contributes to the relativistic degrees of freedom. In the opposite regime of dark matter particles with $m_\chi >T_{\text{DS,*}}$, the dark matter is typically subdominant to dark radiation right after the Dark Big Bang (since heavy particles are energetically difficult to produce), such that $g_{\text{DS}}(T_{\text{DS,*}})$ is well approximated by the number of dark radiation species. Note that $T_{\text{DS},*}$ is generically different from the visible sector temperature $T_*$ since the two sectors are decoupled from each other.

\subsection{Strength of the Dark Big Bang}\label{sec:strength}

The strength of the Dark Big Bang can be measured by the parameter $\alpha$ which is defined as~\cite{Kamionkowski:1993fg}
\begin{equation}\label{eq:alpha}
\alpha = \frac{\rho_\phi}{\rho_{r,*}}\,,
\end{equation}
where we introduced $\rho_{r,*}=\rho_r(T_*)$, i.e., the radiation density of the visible sector just after the Dark Big Bang. The larger $\alpha$ is chosen, the more energy density participates in the phase transition and, hence, the stronger the Dark Big Bang. 

The value of $\alpha$ also determines the ratio of dark-to-visible-sector temperature right after the Dark Big Bang,
\begin{equation}\label{eq:Tstratio}
\frac{T_{\text{DS,*}}}{T_*} = \alpha^{1/4} \left(\frac{g_{\text{eff}}(T_*)}{g_{\text{DS}}(T_{\text{DS,*}})}\right)^{1/4}\,,
\end{equation}
where we employed Eq.~\eqref{eq:darkreheating}.

We first derive a lower limit on $\alpha$. For this purpose, we require that the Dark Big Bang accounts for the entire dark matter density $\rho_{\text{DM},0}$ in today's Universe. This translates to a constraint on the vacuum energy liberated in the phase transition,
\begin{equation}\label{eq:constraintrhophi}
\rho_{\phi} \geq \frac{\rho_{\text{DM},0}}{a_*^3}\,,
\end{equation}
where $a_*$ denotes the scale factor at the Dark Big Bang. The limit above is saturated if all vacuum energy is immediately transferred to decoupled non-relativistic dark matter. If additional dark radiation was produced in the phase transition, if the dark matter was initially relativistic or if dark matter number changing reactions were active for some time, this would only make the constraint stronger. Imposing Eq.~\eqref{eq:constraintrhophi} leads to the following lower bound on $\alpha$,
\begin{equation}\label{eq:alphamin1}
\alpha \geq
\frac{\rho_{\text{DM},0}}{ a_*^3\,\rho_{r,*}} =  \frac{4}{3}\frac{\rho_{\text{DM},0}}{s_0\,T_*}=5.8\times 10^{-4} \left( \frac{\text{keV}}{T_*}\right)\,,
\end{equation}
where we expressed the radiation density in terms of the visible-sector entropy density at the Dark Big Bang $s_*=(4/3)\,\rho_{r,*}/T_*$ and applied entropy conservation $s_* a_*^3 = s_0$. In the last step we plugged in the observed dark matter density $\rho_{\text{DM},0}=1.26\:\text{keV}\text{cm}^{-3}$ as well as today's entropy density $s_0=2890\:\text{cm}^{-3}$~\cite{Planck:2018vyg}.

In the remainder of Sec.~\ref{sec:strength}, we will find the maximal strength of the Dark Big Bang, i.e. we will obtain an upper limit on $\alpha$. The presence of extra energy density in the form of $\rho_{\text{DS}}$ increases the Hubble expansion rate $H$. This so-called ``speed-up effect'' can cause weak interactions to freeze out at higher temperature. As a consequence a larger neutron-to-proton ratio arises at the beginning of BBN which causes an increase of the $^4\text{He}$-fraction as compared to standard cosmology~\cite{Steigman:1977kc}. In addition, the speed-up effect also impacts the CMB by reducing the power in its damping tail (see e.g.~\cite{Hou:2011ec}). Hence, the measured light element abundances (in particular $^4\text{He}$) and the small-scale CMB power spectrum can be used to set upper limits on the extra energy density which are typically presented as constraints on the effective number of extra neutrino species $\Delta N_{\text{eff}}$,
\begin{equation}\label{eq:deltaneffth}
\Delta N_{\text{eff}}(T) \simeq 3\,\frac{\rho_{\text{DS}}(T)-\rho_{\text{DM},0}\,a^3(T)}{\rho_\nu(T)}\,,
\end{equation} 
where $\rho_\nu$ is the energy density in the three active neutrino species, and
$\rho_{\text{DM},0}\,a^3$ the dark matter energy density of $\Lambda$CDM (which needs to be subtracted in the above expression such that $\Delta N_{\text{eff}}(T)=0$ for a dark sector containing only the cold dark matter relic). Notice that $\Delta N_{\text{eff}}$ is in general temperature-dependent since $\rho_{\text{DS}}$ and $\rho_\nu$ may redshift differently.

The Planck collaboration has obtained $\Delta N_{\text{eff}}< 0.3$ at $95\%$ confidence level (CL) by combining CMB and BAO data~\cite{Planck:2018vyg}. However, if local measurements of the Hubble parameter $H_0$~\cite{Riess:2018uxu} are also included, the fit yields $\Delta N_{\text{eff}} = 0.22 \pm 0.15$~\cite{Planck:2018vyg} which suggests
\begin{equation}\label{eq:deltaneff}
\Delta N_{\text{eff}} < 0.5\,,
\end{equation}
at $95\%$ CL. A small positive $\Delta N_{\text{eff}}$ is preferred since it somewhat eases the Hubble tension -- the discrepancy between local and CMB measurement of $H_0$. In the following we shall employ the more conservative (weaker) constraint Eq.~\eqref{eq:deltaneff} which translates to the following bound on the dark radiation density in the present Universe (the bound is on the sum of dark radiation plus an additional possible gravitational wave component) 
\begin{equation}\label{eq:rhodr0}
\rho_{\text{DR},0} = \Delta N_{\text{eff}} \,\frac{7}{4} \,\frac{\pi^2}{30} T_{\nu,0}^4 \quad\Longrightarrow\quad \rho_{\text{DR},0,\text{max}}
=29.6\:\text{meV}\text{cm}^{-3}\,,
\end{equation}
where we used $T_{\nu,0} =1.95\:\text{K}$. In contrast to the CMB limit above, BBN constraints exhibit a stronger model-dependence. This is because the induced element abundances are sensitive to $\Delta N_{\text{eff}}$ at the time of BBN. The latter does, however, map to very different values of $\rho_{\text{DR},0}$ depending on whether the Dark Big Bang occurred prior to or after BBN (which determines how much the dark sector energy density has redshifted by today). Luckily, we can ignore this subtlety, since the BBN constraints are either comparable to or weaker than the CMB limit.\footnote{CMB constraints are superior to BBN constraints if $\Delta N_{\text{eff}}(T)$ increases or remains constant between the BBN and the CMB epoch. This condition is satisfied in the Dark Big Bang scenario.
}
Hence it is sufficient to apply Eq.~\eqref{eq:rhodr0} in the following.

The maximal dark sector energy density today is given by $\rho_{\text{DS},0,\text{max}}=\rho_{\text{DM},0}+\rho_{\text{DR},0,\text{max}}$ with $\rho_{\text{DR},0,\text{max}}$ from Eq.~\eqref{eq:rhodr0}.
But in order to derive the maximal dark sector energy density at the Dark Big Bang $\rho_{\text{DS},*,\text{max}}$ we also need to account for redshifting. In order to maximize $\rho_{\text{DS},*}$, we assume the largest possible redshift between the Dark Big Bang and today. While $\rho_{\text{DR}}\propto a^{-4}$ in the entire post-Dark-Big-Bang evolution, structure formation requires the dark matter to behave as a cold relic at temperatures of the Universe below a few keV (see Sec.~\ref{sec:wdm}) corresponding to scale factors $a>a_{\text{nr}}\sim 10^{-7}$, where $a_{\text{nr}}$ is the value of the scale factor at the time the dark matter particles must become non-relativistic. Hence, we arrive at
\begin{equation}\label{eq:rhophimax1}
\rho_{\phi,\text{max}} = \rho_{\text{DS},*,\text{max}} = \rho_{\text{DR},0,\text{max}}\;\frac{1}{a_*^4} + \rho_{\text{DM},0}\;\frac{\max(a_{\text{nr}},a_*)}{a_*^4}\simeq \frac{\rho_{\text{DR},0}}{a_*^4}\,.
\end{equation}
In the last step we neglected the contribution proportional to the dark matter density. We will show in Sec.~\ref{sec:adiabatic} that structure formation also imposes a constraint $a_* \lesssim 10^{-7}$ on the scale factor of the Dark Big Bang. Therefore, $\rho_{\text{DM},0}\max(a_{\text{nr}},a_*)\ll \rho_{\text{DR},0}$ which justifies the omission of the subleading term.

\begin{figure}[htp]
\begin{center}
\includegraphics[width=0.65\textwidth]{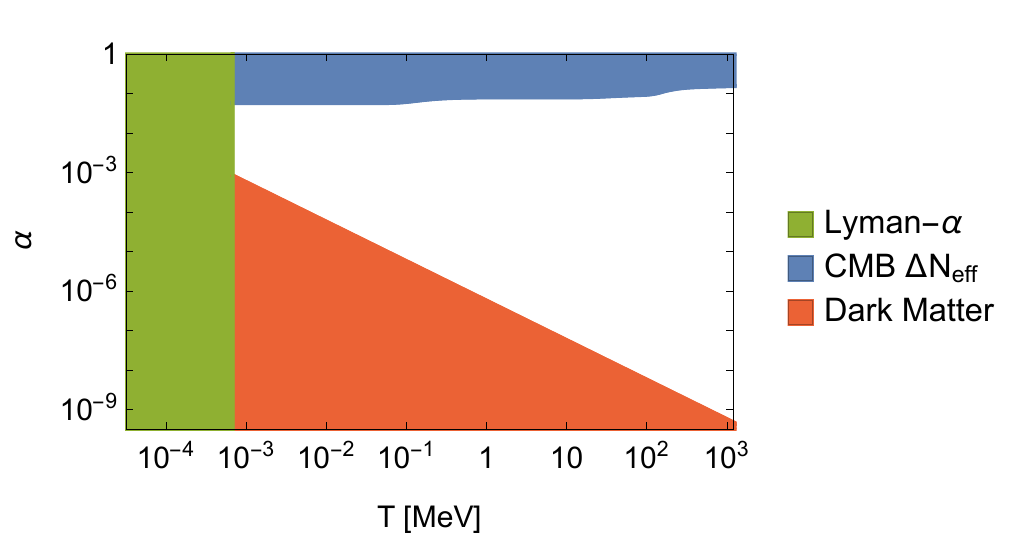}
\end{center}
\caption{Allowed strength $\alpha$ of the Dark Big Bang phase transition as a function of the (visible sector) temperature. The colored regions are excluded by Ly-$\alpha$ constraints on structure formation, CMB bounds on $\Delta N_{\text{eff}}$ and underproduction of dark matter respectively. At higher temperatures (beyond those shown in the picture), the dark-matter bound (red) continues to become weaker, while the $\Delta N_{\text{eff}}$ bound (blue) changes only slightly.}
\label{fig:alpharange}
\end{figure}

We can now combine Eq.~\eqref{eq:rhodr0} and Eq.~\eqref{eq:rhophimax1} to obtain an upper limit on the strength of the Dark Big Bang,
\begin{equation}\label{eq:alphamax}
\alpha < \frac{\rho_{\text{DR},0}}{\rho_{r,*}\,a_*^4}=\left(\frac{2\pi^2}{45}\right)^{1/3}\frac{4\,g^{1/3}_{\text{eff}}(T_*)\,\rho_{\text{DR},0}}{3\,s_0^{4/3}}<0.079 \left(\frac{g_{\text{eff}}(T_*)}{10}\right)^{1/3}
\end{equation}
Eq.~\eqref{eq:alphamin1} and Eq.~\eqref{eq:alphamax} define the minimal and maximal strength of the Dark Big Bang. 

We depict the allowed range of $\alpha$ as a function of the visible sector temperature at the Dark Big Bang in Fig.~\ref{fig:alpharange}. For a late Dark Big Bang around or after the BBN epoch ($T_*\lesssim 10\:\text{MeV}$) we find,
\begin{equation}\label{eq:alpharange}
6\times 10^{-8} < \alpha < 0.08 \,\,\,\,\,\,\,\,\,\,\,\,\,\, ({\rm for} \, T_*\lesssim 10\:\text{MeV}) \, .
\end{equation}
As shown above and in Fig.~\ref{fig:alpharange}, the lower bound is from the requirement of producing enough dark matter, while the upper bound is from constraints on $\Delta N_{\rm eff}$ from the CMB.

Since $T_{\text{DS},*}\sim \alpha^{1/4} T_*$ (cf.~Eq.~\eqref{eq:Tstratio}), the dark sector temperature at the Dark Big Bang is usually lower than the visible sector temperature.\footnote{
Since the dark sector energy density can make up at most $\sim 10\%$ of the total energy density at the Dark Big Bang (see our constraints on $\alpha$ above) it follows that $T_{\text{DS},*}$ is almost always smaller than $T_*$ (cf.\ Eq.~\eqref{eq:Tstratio}). An exception occurs if (i) the dark sector density saturates its upper limit and (ii) the dark sector contains much fewer degrees of freedom compared to the visible sector. In this extreme case $T_{\text{DS},*}$ can be up to a factor of~2 larger than $T_*$.} For a late Dark Big Bang around or after BBN the difference is, however, at most two orders of magnitude. If, on the other hand, the Dark Big Bang occurs long before BBN ($T_*\gg \text{MeV}$), the lower bound on $\alpha$ becomes considerably weaker and a stronger suppression of $T_{\text{DS},*}/T_*$ may arise.

\subsection{Comment on Initial Conditions and Inflation Scale}\label{sec:initial}

The scenario we described begins with the observable Universe being trapped in the metastable minimum of $\phi$. This situation is realized through inflation which blows up an initial patch of dark false vacuum to contain the entire observable Universe\footnote{The simplest possibility is that there were random false and true vacuum patches before inflation of which one false vacuum patch was blown up to contain the entire observable Universe.}.
However, in order not to destabilize the dark false vacuum already during inflation through fluctuations caused by the Gibbons-Hawking temperature $T_{\text{GH}}\sim H_I$~\cite{Gibbons:1977mu}, where $H_I$ is the Hubble constant during inflation, a constraint on the scale of inflation arises. Specifically we need to require that $T_{\text{GH}}^4$ does not exceed the barrier height $V_b$ protecting the metastable minimum~\cite{Hawking:1981fz}. The potential barrier itself is bounded by $V_b \lesssim \Delta V$ since the false vacuum would otherwise become stable within the lifetime of the Universe.\footnote{The lifetime of the false vacuum increases exponentially with the barrier height $V_b$.} From Eq.~\eqref{eq:alphamax} it, furthermore, follows that $ (\Delta V)^{1/4}\lesssim T_*$. Stability of the false vacuum during inflation thus requires at least $H_I\lesssim T_*$ which translates to an inflation scale,
\begin{equation}\label{eq:VI}
V_I^{1/4} \lesssim 10^8\gev\times \sqrt{\frac{T_*}{\text{MeV}}}\,.
\end{equation}
If we consider a phase transition around BBN as an example, Eq.~\eqref{eq:VI} would impose a small inflation scale $V_I^{1/4} \lesssim 10^8\gev$ (while an earlier Dark Big Bang would allow for a somewhat larger inflation scale). Such low-scale inflation has recently become popular due to the Trans-Planckian Censorship Conjecture~\cite{Bedroya:2019snp} which postulates that inflation with $V_I^{1/4}>10^9\gev$ cannot arise in a consistent quantum theory of gravity~\cite{Bedroya:2019tba}. Therefore, low-scale inflation constitutes a simple and well-motivated history supporting a Dark Big Bang cosmology.

However, an inflation scale constrained by the upper limit in Eq.~\eqref{eq:VI} is far from the only possibility to set the right initial conditions for a Dark Big Bang.
In fact, for a larger inflation scale some patches of the Universe may still fall into the false vacuum after inflation once the Hubble scale drops below $V_b^{1/4}$. Patches of true and false vacuum would be separated by domain walls in this case.  Successful dark matter production could then be realized from the annihilation of the domain walls~\cite{Vilenkin:1981zs,Gelmini:1988sf} which would play the role of the Dark Big Bang.

Alternatively, even if $\phi$ is displaced from its false vacuum during inflation, thermal corrections may push the entire Universe into the dark false vacuum after the Hot Big Bang. While we assumed that the Hot Big Bang dominantly heats the visible sector (which does not couple to $\phi$) even very subdominant dark radiation plasma can initially stabilize the dark false vacuum (see e.g.~\cite{Jaeckel:2016jlh,Addazi:2017gpt,Breitbach:2018ddu,Niedermann:2021ijp}). Yet another possibility consistent with high-scale inflation invokes a second dark sector scalar field, coupled to $\phi$, that leads to a change in the tunneling rate over time.  For example, the evolution of the second scalar field can
allow for a dynamical barrier height which is initially large -- thus strongly stabilizing $\phi$ in the dark false vacuum during inflation -- but subsequently decreases to allow for the false vacuum to decay within the age of the Universe (similar to the time-changing tunneling rate proposed in double field inflation~\cite{Adams:1990ds,Linde:1990gz} and new early dark energy~\cite{Niedermann:2019olb}).

We conclude that many plausible cosmological histories exist in which the Universe is (or patches of the Universe are) trapped in the false vacuum during the early radiation-dominated epoch. The realization of a Dark Big Bang thus emerges as a very natural possibility.

\section{Constraints from Structure Formation and the CMB}\label{sec:structure}

In this section we will investigate the evolution of perturbations in the dark sector. Specifically, we will examine isocurvature,  bubble-induced and adiabatic perturbations. We will first argue that potentially dangerous isocurvature perturbations, which are seeded during inflation, are quickly redshifted away and do not impact the late-time evolution of the Universe. Then, we will impose that perturbations induced by the bubble collisions during the phase transition occur on sufficiently small scales such that they do not spoil CMB and Large Scale Structure observations. Finally, we will require that -- on cosmologically observable scales -- the dark matter fluid picks up and sustains the right adiabatic perturbations to support successful structure formation. This will allow us to constrain the time of the Dark Big Bang and the phase space properties of dark matter.

\subsection{Isocurvature Perturbations}\label{sec:isocurvature}

Since the origin of radiation, baryons (Hot Big Bang) and dark matter (Dark Big Bang) are different in our scenario, one might worry that dark matter receives dangerous isocurvature perturbations which are strongly constrained by observation. A possible source of isocurvature perturbations are fluctuations of the tunneling field,
\begin{equation}
\delta \phi = \phi - \bar{\phi}\,,
\end{equation}
where $\bar{\phi}$ denotes the mean field value. Such fluctuations can be generated randomly by quantum jumps of $\phi$ during inflation (see e.g.~\cite{Linde:2005yw}). Because $\delta \phi$ is uncorrelated with the inflaton fluctuation, it amounts to a potentially dangerous isocurvature perturbation~\cite{Linde:1985yf,Kofman:1985zx,Kofman:1986wm} which is inherited by the dark matter component at the Dark Big Bang. However, we will argue that isocurvature perturbations are (virtually) never a problem because $\delta \phi$ is extremely tiny by the time of the Dark Big Bang. 

Depending on the mass of the tunnelling field $m$ in the false vacuum, we can distinguish two cases: 
\begin{enumerate}
\item $m > H_I\,$,
\item $m \lesssim H_I$,
\end{enumerate} 
where $H_I$ is the Hubble constant during inflation (as above).
In case~1 quantum fluctuations of $\phi$ during inflation are subject to a strong suppression such that isocurvature perturbations are never generated. 
In case~2, the tunneling field acquires a spectrum of perturbations during inflation~\cite{Bunch:1978yq}
\begin{equation}
\delta\phi_k \sim \frac{H}{2\pi}\,,
\end{equation}
where we introduced the Fourier modes of the field fluctuation $\delta\phi_k$. The subsequent evolution of $\delta\phi_k$ follows from the equation of motion (see e.g.~\cite{Hwang:1993cv})\footnote{We assume that the minimum of $\phi$ during and after inflation coincide. If this is not the case, an additional term $\propto \dot{\bar{\phi}}$ would be present in the equation of motion during the radiation-dominated stage. Such a term would, however, typically redshift away quickly and, thus, not change the conclusions of this section.},
\begin{equation}\label{eq:perturbationeq}
\ddot{\delta\phi}_k + 3 H \dot{\delta\phi}_k + m^2\,\delta\phi_k + \frac{k^2}{a^2}\delta\phi_k = 0\,.
\end{equation}
Once, a scale has exited the horizon during inflation, the last term on the left-hand-side becomes negligible (until horizon re-entry). In the super-horizon regime $\delta\phi_k$ thus obeys the homogeneous Klein-Gordon equation and behaves as a classical scalar field in an expanding Universe.

During the slow-roll regime of inflation, the first term on the left hand side of Eq.(\ref{eq:perturbationeq}) is negligible 
so that the solution to the equation becomes (see e.g.~\cite{Lemoine:2009is})
\begin{equation}
\label{eq:soln}
\delta\phi_k \propto  \exp\left( -\frac{m^2}{3 H_I^2} N_k \right)\qquad\text{(during inflation)}\,,
\end{equation}
where $N_k$ denotes the number of e-folds after horizon exit of the scale $k$.
Here $N_k\sim 30-60$ for the scales we observe in the CMB and in Large Scale Structure.
We can now distinguish two subcases: if $m\sim H_I$, the perturbation $\delta\phi_k$ on observable scales is washed out during inflation, due to the strong exponential suppression of $\delta\phi_k$ on the right hand side of Eq.~\eqref{eq:soln}. Thus, isocurvature perturbations (on relevant scales) do not survive the inflationary epoch in this case.

On the other hand, if $m\ll H_I$, the perturbation $\delta\phi_k$ is frozen during inflation. In the subsequent radiation-dominated epoch $\delta\phi_k$ remains approximately constant until $H\sim m$. But once the Hubble rate drops below the mass, $\delta\phi$ commences oscillations around $\delta\phi=0$ which are damped by the Hubble friction~\cite{Moroi:2001ct}. In the oscillatory regime, the mean squared field value $\langle \delta\phi_k^2\rangle$ -- which sets the isocurvature perturbation in the density $\rho_\phi$ -- redshifts as
\begin{equation}
\langle \delta\phi_k^2\rangle\propto \frac{1}{a^3}\,.
\end{equation}
This is in contrast to the mean energy density $\bar{\rho}_\phi$ which remains constant (since vacuum energy does not redshift). Therefore, once the oscillations set in, the isocurvature perturbation quickly redshifts away. In the absence of any small couplings, the Dark Big Bang scenario typically features $m\sim(\Delta V)^{1/4} \sim \sqrt{H_* M_P} \gg H_*$ such that the oscillation stage continues for a significant time period. By the time of the Dark Big Bang $\langle \delta\phi_k^2\rangle$ is so tiny that it induces no measurable perturbation in the dark matter component. Unless for extreme parameter choices, we can thus safely neglect isocurvature perturbations from inflation. This conclusion holds independent of the initial conditions leading to a metastable false vacuum in the early Universe (see Sec.~\ref{sec:initial}).

\subsection{Perturbations from Bubble Collisions}\label{sec:bubble}

Another source of perturbations in the dark matter plasma are the bubble collisions. These induce non-linearities whose physical size is controlled by the radius of the colliding bubbles $R_b$ (see e.g.~\cite{Niedermann:2020dwg,Freese:2021rjq}). Since bubbles approximately expand at the speed of light we can estimate $R_b$ by the typical distance between the nucleation sites,
\begin{equation}\label{eq:Rb}
R_b \simeq \Gamma^{-1/4}\,.
\end{equation}
This approximation neglects the expansion of the Universe between bubble nucleation and bubble collision. The corresponding error is, however, negligible since the phase transition completes within a fraction of a Hubble time (cf. Eq.~\eqref{eq:beta}). The comoving scale $k_b$ corresponding to the size $R_b$ at the time of the Dark Big Bang reads
\begin{equation}
k_b \simeq \frac{\pi}{R_b/a_*} \simeq \frac{2}{\text{Mpc}}\left(\frac{10^{-5}}{a_*}\right)\,,
\end{equation}
where we used Eq.~\eqref{eq:Rb}, Eq.~\eqref{eq:tstar} and the time-scale-factor relation of radiation-domination in order to express $R_b$ in terms of the scale factor of the Dark Big Bang $a_*$.

The bubble collision, hence, induce a feature in the spectrum of dark matter perturbations which is peaked at $k_b$. The absence of such a feature in the CMB and in the matter power spectrum allows us to constrain the time of the Dark Big Bang. Specifically, we need to require that $k_b$ falls outside the range of scales which are accessible either by CMB~\cite{Planck:2018vyg} observations or by measurements of the Lyman-$\alpha$ (Ly-$\alpha$) absorption in distant quasars (see e.g.~\cite{Croft:1997jf,Viel:2005qj,Viel:2013fqw}),
\begin{equation}\label{eq:krange}
k_{\text{CMB}}\simeq (10^{-4}-0.5)\:\text{Mpc}^{-1}\,, \qquad
k_{\text{Ly-}\alpha}\simeq (0.1-10)\:h\text{Mpc}^{-1}\,.
\end{equation}
We note that Ly-$\alpha$ data cover a regime of scales in which cosmological perturbations are already affected by non-linear evolution. The resulting mixing of scales would somewhat wash out the peak in the matter power spectrum at the scale of the colliding bubbles. Nevertheless, due to the prominence of the feature, Ly-$\alpha$ data exclude $k_b$ in the range of $k_{\text{Ly-}\alpha}$ indicated above. 

The strongest constraint on the Dark Big Bang is set by the largest $k$ (smallest observable scale). While $k_{\text{CMB}}$ is limited by the angular resolution of the instruments, the upper limit on $k_{\text{Ly-}\alpha}$ is imposed by the contamination induced by metal lines and by the effective Jeans scales below which pressure gradients wipe out small scale fluctuations in the baryons. From $k_b > 10\:h\text{Mpc}^{-1}$ we obtain the following constraint on the scale factor of the Dark Big Bang,
\begin{equation}\label{eq:bubblebound}
a_* < 3\times 10^{-6}\,.
\end{equation}
In order not to spoil Ly-$\alpha$ observations by the anisotropies generated through bubble collisions, the Dark Big Bang should, hence, have happened at most 6~years after the Hot Big Bang. This corresponds to a temperature and redshift\footnote{This bound holds for a Dark Big Bang phase transition with a constant $\Gamma$. It can be circumvented in some scenarios with a time-dependent $\Gamma$~\cite{Adams:1990ds,Niedermann:2019olb}.}  
\begin{equation}
T_*>80\:\text{eV},  \,\,\,\,\,\,\,\,\,\,\,\,\,\,\,\,\,   z_*>3\times 10^5 \, .
\end{equation}
Here we can already see that a Dark Big Bang first-order phase transition cannot take place as late as the epoch of matter-radiation equality at $z=3500$ (or as late as the epoch of Early Dark Energy shortly before matter-radiation equality). We will, however, see in the next section that a tighter constraint can be derived by requiring the right adiabatic dark matter perturbations.

So far, we have approximated the time of the true-vacuum bubble nucleation by $t_*$. This appears justified since most bubbles are generated in a narrow time window of order $\beta^{-1}$ around $t_*$. However, we need to take into account that a large number of bubbles was nucleated in our past Universe. Therefore, one might worry about statistical outliers: some rare bubbles which appeared significantly before $t_*$. These early-produced bubbles would have had substantially more time to grow compared to the average bubbles nucleated around $t_*$. Even a single such ``big bubble'' could potentially have disastrous effects by generating a large-scale anisotropy in the Universe~\cite{Weinberg:1989mp,La:1989pn,La:1990wy,Liddle:1991tr,Turner:1992tz}. This issue was originally raised in the context of a phase transition at the end of inflation and is referred to as the ``big bubble problem''. Luckily, it turns out that big bubbles do not pose a threat in the Dark Big Bang scenario. This can be understood by considering the radius $R_{\text{big}}$ of a big bubble emitted at the time $t_{\text{big}}$,
\begin{equation}\label{eq:rbig}
 R_{\text{big}}(t) = a(t) \int\limits_{t_{\text{big}}}^t \frac{dt^\prime}{a(t^\prime)}\,,
\end{equation}
where we assumed the bubble to expand at the speed of light. We will now consider a big bubble nucleated (i) during radiation domination (but significantly before $t_*$) (ii) during inflation.

If the big bubble was nucleated during radiation domination we can set $a(t)\propto t^{1/2}$ in which case Eq.~\eqref{eq:rbig} immediately implies $R_{\text{big}}(t)\leq 2\,t$.\footnote{The bound $R_{\text{big}}(t)\leq 2\,t=H^{-1}$ also trivially follows from the argument that physical scales cannot exit the horizon during a stage of subluminal expansion.} At the time $t_*$, the big bubble would collide with average bubbles of size $R_b\sim \Gamma^{-1/4}$. By then the big bubble has grown to a size $R_{\text{big}}(t_*)\leq 3\,\Gamma^{-1/4}$, where we used Eq.~\eqref{eq:tstar}. Hence, we see that the big bubble size at collision is at most an $\mathcal{O}(1)$-factor larger than the average bubble size. Hence, big bubbles generated during radiation domination do not considerably alter the bound in Eq.~\eqref{eq:bubblebound}.

The situation is somewhat different for big bubbles of the dark phase transition which are already nucleated during inflation. Due to the superluminal expansion such bubbles can grow exponentially during inflation $R_{\text{big}}\propto e^{N_\text{big}}$, where $N_\text{big}$ denotes the number of e-folds before the end of inflation when the big bubble was emitted. Luckily, however, the probability that even a single bubble was nucleated within our observable Universe during inflation is extremely low in the Dark Big Bang scenario. In~\cite{Turner:1992tz} the number of big bubbles (generated during inflation) emerging on the last scattering surface was estimated as $N_{\text{big}}\sim 10^4 \times \Gamma/H_I^4$, where $H_I$ is the Hubble scale during inflation. While this estimate was performed for a (visible sector) phase transition at the end of inflation in~\cite{Turner:1992tz} it should also apply to the Dark Big Bang phase transition. Since $\Gamma\sim 10^2 H_*^4$ in the Dark Big Bang scenario (cf.\ Eq.~\eqref{eq:tstar}), we see that big bubble formation during inflation is completely negligible if the Hubble scale of the Dark Big Bang is two or more orders of magnitude smaller than the Hubble scale during inflation. This is a very mild constraint since we usually expect the inflation scale to be many orders of magnitude larger than the Dark Big Bang energy scale. We can conclude that -- unless in some pathological cases -- big bubble formation is never a problem in the Dark Big Bang scenario.

\subsection{Adiabatic Perturbations}\label{sec:adiabatic}

The evasion of unwanted isocurvature/ peaked perturbations does not suffice for a successful cosmological scenario. We also need to require that dark matter produced by the Dark Big Bang obtains the adiabatic perturbations which are the seed for structure formation. Since the dark sector is decoupled from the SM radiation bath, it might seem that dark matter does not receive any adiabatic fluctuations at all. However, we will show that even in the absence of direct couplings between dark and visible matter, gravity will imprint the fluctuations in the radiation bath onto dark matter at the Dark Big Bang. 

We first consider the density fluctuation of the SM radiation bath $\delta\rho_r=\rho_r - \bar{\rho}_r$ and the fluctuation in the Hubble scale $\delta H = H -\bar{H}$ neglecting the impact of the dark sector. Quantities with a bar refer to the mean quantities (averaged over all patches in the Universe). Our assumption is justified during the radiation-dominated epoch, where $\rho_{\text{DS}}$ is subdominant. The evolution equations (in the comoving gauge) are thus~\cite{Lyth:1988pj},
\begin{align}\label{eq:deltaH}
\delta \dot{\rho}_r &= - 4\rho_r \delta H - 3 H \delta\rho_r\,, \nonumber\\
\delta \dot{H} &= -2H \delta H - \frac{1}{6} \delta\rho - \frac{\nabla^2 \delta P}{12 \rho_r}\,.
\end{align}
In the super-horizon regime the pressure gradient term (last term on the right-hand-side) can be neglected. The fluctuations quickly reach their asymptotic solutions which, in terms of Fourier modes, are expressed as~\cite{Lyth:1988pj}
\begin{equation}\label{eq:drhosolution}
\frac{\delta \rho_{r,k}}{\rho_r} = \frac{4}{9}\,\left(\frac{k}{aH}\right)^2 \mathcal{R}_k\,,\qquad
 \frac{\delta H_k}{H} = -\frac{1}{9}\left(\frac{k}{aH}\right)^2 \mathcal{R}_k\,,
\end{equation}
where $\mathcal{R}_k$ is constant and measures the perturbation in the spatial curvature of comoving hypersurfaces~\cite{Bardeen:1980kt}.

Now we turn to the perturbations in the dark matter component. At the phase transition, the bubble collisions generate fluctuations which are imprinted onto the dark matter density $\rho_\chi$. However, these are strongly peaked at the comoving length scale $1/k_b$ which must be small enough not to spoil CMB and Ly-$\alpha$ observations (see Sec.~\ref{sec:bubble}). As long as we are only interested in larger cosmological scales (namely those accessible in the CMB and Ly-$\alpha$), we can ``integrate out'' the peaked perturbations~\cite{Baumann:2010tm}.

Let us also point out that $\rho_\phi$ is effectively smooth at the time of the Dark Big Bang. This is because isocurvature perturbations of $\rho_\phi$ -- if they are generated during inflation -- do usually not survive until $t_*$ (see Sec.~\ref{sec:isocurvature}). Furthermore, $\rho_\phi$ does not pick up the adiabatic perturbations of the radiation bath since vacuum energy does not support its own perturbation. Hence, dark matter cannot inherit any perturbations from $\phi$. It may, therefore, naively seem as if dark matter was produced as an effectively smooth fluid by the Dark Big Bang. However, this is not the case. We will argue in the following that the radiation bath imprints its fluctuation onto the dark matter by modulating the time of the Dark Big Bang.

When the dark false vacuum decays, the Universe is still dominated by the radiation bath which exhibits adiabatic perturbations. As is well-known from cosmological perturbation theory, adiabatic perturbations have the property that local quantities like energy density and pressure at some spacetime point in the perturbed Universe
are the same as those in the background Universe at a slightly different time (see e.g.~\cite{Baumann:2018muz}). Hence,
\begin{equation}
\delta\rho_r= \rho_r(t+\delta t)-\rho_r(t) = \dot{\rho}_r \delta t\,,
\end{equation}
where we note that $\delta t$ depends on the space-coordinate. Adiabatic perturbations thus imply that some patches of the Universe are ahead and others behind in the evolution with the time difference,
\begin{equation}\label{eq:deltat}
\delta t = \frac{\delta \rho_r}{\dot{\rho}_r}= -\frac{\delta \rho_r}{4H\rho_r}\,.
\end{equation}
As a consequence, different regions of the Universe will undergo the Dark Big Bang phase transition at slightly different times (see~\cite{Cruz:2022oqk}). The earlier the dark matter is produced, the more time it has to redshift (i.e.\ reduce its energy density). Therefore, a positive $\delta t(t_*)\equiv \delta t_*$ causes a local underdensity in the dark matter fluid. We can derive the perturbed dark matter density at the Dark Big Bang by taking into account the redshift between $t_*-\delta t_*$ and $t_*$,
\begin{equation}\label{eq:rhochi}
\rho_{\chi,*} = \bar{\rho}_{{\chi,*}}\;\exp\left(-3 \int\limits_{t_*-\delta t}^{t_*} dt (1+w) H\right)\,,
\end{equation}
where $w$ denotes the dark matter equation-of-state parameter (which can be time-dependent). In the limiting cases of a highly relativistic and a decoupled non-relativistic species we can set $w=1/3$ and $w=0$ respectively. Expanding Eq.~\eqref{eq:rhochi} in $\delta t$ yields, 
\begin{equation}\label{eq:drhochistar}
\frac{\delta \rho_{\chi,*}}{\rho_{\chi,*}} = - 3(1+w) H_*\delta t_* = \frac{3(1+w)}{4}\frac{\delta \rho_{r,*}}{\rho_{r,*}}\,,
\end{equation}
where we employed Eq.~\eqref{eq:deltat} in the second step. Hence, we find,
\begin{equation}
\frac{\delta \rho_{\chi,*}}{\dot{\rho}_{\chi,*}} = \frac{\delta \rho_{r,*}}{\dot{\rho}_{r,*}}\,,
\end{equation}
which is precisely the condition for an adiabatic perturbation. We can conclude that the radiation bath imprints its adiabatic perturbation onto the dark matter at the Dark Big Bang.

The dark matter perturbation will remain adiabatic subsequently which can be shown by considering the evolution equation of the dark matter perturbation during the radiation-dominated epoch,
\begin{equation}
\delta \dot{\rho}_\chi= - 6(1+w)\rho_\chi\delta H - 3 (1+w) H \delta\rho_\chi\,.
\end{equation}
The solution to the above equation can be written as
\begin{equation}\label{eq:drhochi}
\frac{\delta \rho_{\chi}}{\rho_{\chi}} = \frac{\delta \rho_{\chi,*}}{\rho_{\chi,*}}+
\frac{3(1+w)}{4}\left( \frac{\delta \rho_r}{\rho_r}-\frac{\delta \rho_{r,*}}{\rho_{r,*}}\right)\,,
\end{equation}
where we can neglected the (tiny) backreaction of the dark matter fluctuation onto $\delta H$, $\delta \rho_r$ and employed the solutions from Eq.~\eqref{eq:drhosolution}. Plugging Eq.~\eqref{eq:drhochistar} into Eq.~\eqref{eq:drhochi} yields
\begin{equation}\label{eq:drhochi2}
\frac{\delta \rho_{\chi}}{\rho_{\chi}}  = \frac{3(1+w)}{4}\frac{\delta \rho_r}{\rho_r} \;\;\;\;\Longleftrightarrow\;\;\;\;\frac{\delta \rho_{\chi}}{\dot{\rho}_{\chi}}  = \frac{\delta \rho_r}{\dot{\rho}_r}\,,
\end{equation}
which proves that the dark matter perturbation remains adiabatic.

Let us make an interesting side remark: we have shown that the Dark Big Bang directly imprints the desired adiabatic perturbation onto the dark matter. However, even if this were not the case and dark matter were produced as an entirely smooth fluid, it would quickly pick up the perturbations of the radiation bath through gravitational interactions. This immediately follows from Eq.~\eqref{eq:drhochi} after taking into account that  the first term inside the parentheses on the right hand side of the equation grows in time as $\delta\rho_r/\rho_r \propto t$ (and dominates over the second term as time goes on). As a consequence the dark matter fluctuation would asymptotically approach the solution in Eq.~\eqref{eq:drhochi2} even if $\delta\rho_{\chi,*}$ were zero. As we have seen, this gravitational transmission of fluctuations is not relevant for the Dark Big Bang scenario, where adiabatic dark matter perturbations are generated immediately at the Dark Big Bang. But it plays an important role for other dark matter candidates, for instance for axion dark matter.

\begin{figure}[htp]
\begin{center}
\includegraphics[width=0.6\textwidth]{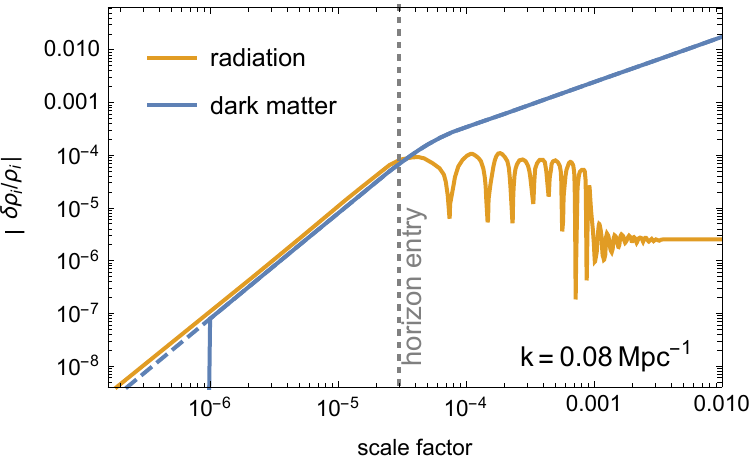}
\end{center}
\caption{Radiation and dark matter perturbations in the Dark Big Bang scenario. A Dark Big Bang at $a=10^{-6}$ is assumed. The dark matter fluid immediately picks up the perturbation of the radiation bath at the Dark Big Bang. Therefore, after the Dark Big Bang, the dark matter perturbation is indistinguishable from the perturbation in $\Lambda$CDM (indicated by the dashed line).}
\label{fig:perturbations}
\end{figure}

In Fig.~\ref{fig:perturbations} we depict the evolution of the radiation perturbation and the dark matter perturbation in the Dark Big Bang scenario.\footnote{The evolution of perturbations after horizon entry was taken from~\cite{Longair:2008gba}.} For illustration we picked a comoving scale of $k=0.08\:\text{Mpc}^{-1}$. The dark matter perturbation vanishes before the Dark Big Bang (at $a_*=10^{-6}$ in the figure), but is immediately generated at $a_*$ due to the fluctuation of the false vacuum decay time inflicted by the radiation perturbation (cf. Eq.~\eqref{eq:deltat}). Notice that in reality the fluctuation is not generated instantaneously, but over the (very short) duration of the Dark Big Bang which we neglected in the figure. At $a>a_*$ the perturbation $\delta\rho_\chi/\rho_\chi$ is indistinguishable from the one in $\Lambda$CDM which is also shown in the figure.

However, it is crucial that dark matter exists and exhibits the correct adiabatic perturbations at the time when observable scales enter the horizon. As can be seen in Fig.~\ref{fig:perturbations}, the evolutions of $\delta\rho_\chi/\rho_\chi$ and $\delta\rho_r/\rho_r$ decouple at horizon entry, when acoustic oscillations in the radiation plasma set in and free-streaming/ diffusion start to affect $\delta\rho_r$. Therefore, if dark matter is only produced after horizon entry of a scale $k$, the fluctuation $\delta\rho_\chi$ imprinted onto the dark matter by the radiation bath does not reflect the in-horizon evolution of $\delta\rho_\chi$ predicted in $\Lambda$CDM at that scale. The Dark Big Bang will thus induce differences in the perturbation spectrum compared to $\Lambda$CDM. In the case of a late Dark Big Bang these occur on observable scales (which are already inside the horizon at the Dark Big Bang) and, therefore, potentially upset structure formation. 

In order to constrain the perturbations in the Dark Big Bang scenario, it is convenient to again employ Ly-$\alpha$ observations of distant quasars which can be used as a tracer of cosmological fluctuations on scales $k_{\text{Ly-}\alpha}\simeq (0.1-10)\:h\text{Mpc}^{-1}$. We would, in principle, need to derive the full linear matter power spectrum in the Dark Big Bang scenario including the case of a late Dark Big Bang. Then, since perturbations in the Ly-$\alpha$ regime are already affected by non-linear evolution, we would need to map the matter power spectrum of the Dark Big Bang scenario onto the flux power spectrum of the quasars -- which involves a complicated dependence on cosmological and astrophysical parameters. 

We decided for a more economic pathway to obtain an approximate bound, where we recast existing constraints from warm dark matter (WDM) simulations.
Specifically we can employ the fact that a late Dark Big Bang suppresses the matter power spectrum at scales $k > a_* H_*$ which are already inside the horizon at $t_*$. This is because the dark matter perturbation generated at the Dark Big Bang reflects the perturbation in the dominant radiation component which does not grow after horizon-entry -- in contrast to the dark matter perturbation of $\Lambda$CDM (see Fig.~\ref{fig:perturbations}). WDM scenarios also induce a suppression of small-scale power -- albeit for a different physical reason. In the case of WDM, the suppression results from the free-streaming of dark matter particles out of small-scale structures. The largest scale affected by free-streaming is the present value of the particle horizon of warm particles which is denoted as the free-streaming horizon $\lambda_{\text{FSH}}$ (see e.g.~\cite{Boyarsky:2008xj}). WDM suppresses the matter power spectrum at $k>\lambda_{\text{FSH}}^{-1}$. Therefore, we expect that the upper limit on the free-streaming horizon $\lambda_{\text{FSH},\text{max}}$ derived in WDM simulations can (approximately) be reinterpreted as the maximal horizon size at $t_*$ in the Dark Big Bang scenario. Hence, in the following, we will require
\begin{equation}\label{eq:astHstmin}
a_* H_* > \lambda_{\text{FSH},\text{max}}^{-1}\,.
\end{equation}
Typically, WDM bounds are presented in terms of the mass $m_{\text{MWD}}$ rather than $\lambda_{\text{FSH}}$ (see e.g.~\cite{Viel:2013fqw,Irsic:2017ixq,Garzilli:2019qki,Villasenor:2022aiy}). However, one can easily translate between the two using the map provided for instance in~\cite{Dienes:2021cxp}. At present, the tightest lower limit\footnote{In order to be conservative we refrain from using the tighter limit $m_{\text{WDM}}>3.1\:\text{keV}$ derived in~\cite{Villasenor:2022aiy} because the same analysis would also exclude $\Lambda$CDM at 95\% CL.} (with minimal assumptions on the thermal history of the Universe) imposes $m_{\text{WDM}}>1.9\:\text{keV}$~\cite{Garzilli:2019qki} which corresponds to~\cite{Dienes:2021cxp}
\begin{equation}\label{eq:LFSH}
\lambda_{\text{FSH},\text{max}} = 0.16\:\text{Mpc}\,.
\end{equation}
After plugging Eq.~\eqref{eq:LFSH} into Eq.~\eqref{eq:astHstmin} and expressing $H_*$ in terms of $a_*$, we obtain,
\begin{equation}\label{astmax}
a_* < 3.4\times 10^{-7}\,.
\end{equation}
Thus, the necessity of adiabatic dark matter perturbations which are sufficiently similar to those in $\Lambda$CDM requires the Dark Big Bang to occur at
\begin{equation}\label{eq:tmaxpert}
t_* < 32.5\:\text{days}\,.
\end{equation}
We note that this bound is only approximate since the Dark Big Bang scenario does not give rise to the precise shape of the power spectrum suppression of WDM. We leave a more in-depth analysis for future studies and, in the following, simply impose Eq.~\eqref{eq:tmaxpert}. We can also translate Eq.~\eqref{eq:tmaxpert} to a minimal (visible-sector) temperature of the Universe at the Dark Big Bang,
\begin{equation}
T_* > 0.68\:\text{keV}\,.
\end{equation}

\subsection{Warm Dark Matter Constraints}\label{sec:wdm}

In the previous section we required the Dark Big Bang to occur early enough to pick up the desired dark matter perturbation on observable scales. However, the constraint of Eq.~\eqref{eq:tmaxpert} can only be saturated if dark matter behaves as a cold relic immediately after the Dark Big Bang. If, on the other hand, dark matter is born relativistic, its free-streaming erases primordial inhomogeneities at scales below the free-streaming horizon $\lambda_{\text{FSH}}$. This provides a complementary mechanism of power suppression which can operate in the Dark Big Bang scenario. Not only must we ensure that the right dark matter perturbation is generated, we also need to require that it persists and is not washed out by the free-streaming.

Since free-streaming can only occur after the Dark Big Bang, we define the free-streaming horizon in the Dark Big Bang scenario in the following way (see e.g.~\cite{Boyarsky:2008xj}),
\begin{equation}\label{eq:lfsh1}
 \lambda_{\text{FSH}}= \int\limits_{t_*}^{t_0}dt\:\frac{\langle v_\chi(t) \rangle}{a(t)}
= \int\limits_{a_*}^{1}da\:\frac{\langle v_\chi(a) \rangle}{a^2H}\,,
\end{equation}
where $t_0$ stands for the present time and $\langle v_\chi(t)\rangle$ for the time-dependent average dark matter velocity. We define $v_*=\langle v_\chi(t_*)\rangle$. Now we distinguish two cases:
\begin{enumerate}
\item dark matter is produced non-relativistically ($v_*\ll 1$),
\item dark matter is produced relativistically ($v_*\simeq 1$).
\end{enumerate}
In case~1 we can employ that the velocity redshifts linearly with the scale factor. Therefore, the free-streaming horizon becomes\,
\begin{equation}\label{eq:lfshnr}
\lambda_{\text{FSH}} \simeq v_*\,a_*\,\int\limits_{a_*}^{1}da\:\frac{1}{a^3H} \simeq
\frac{\sqrt{2}\,a_* v_*}{a_{\text{eq}}^2 H_{\text{eq}}}\log\left(\frac{4 a_{\text{eq}}}{a_{*}}\right)\qquad\text{(non-relativistic)}\,.
\end{equation}
In order to arrive at the above expression, we employed that the integral in Eq.~\eqref{eq:lfsh1} is insensitive to the late-time evolution of the Universe. Therefore we approximated the Hubble rate as $H=H_{\text{eq}} \sqrt{a_{\text{eq}}^3/(2a^3)+a_{\text{eq}}^4/(2a^4)} $, where a subscript `eq' indicates that a quantity is evaluated at matter-radiation equality. Furthermore, we used $a_* \ll a_{\text{eq}} \ll 1$.

In case~2, the dark matter velocity remains approximately constant while being relativistic and redshifts linearly with the scale factor once it becomes non-relativistic. We denote the scale factor of the non-relativistic transition by $a_{\text{nr}}$ and approximate $\langle v_\chi(a) \rangle\simeq 1$ for $a<a_{\text{nr}}$ and $\langle v_\chi(a) \rangle \simeq a_{\text{nr}}/a$ for $a>a_{\text{nr}}$. This allows us to derive the free-streaming horizon for the case of relativistic dark matter production
\begin{equation}\label{eq:lfsh2}
\lambda_{\text{FSH}}
\simeq \int\limits_{a_*}^{a_{\text{nr}}}da\:\frac{1}{a^2H} 
+\int\limits_{a_{\text{nr}}}^{1}da\:\frac{a_{\text{nr}}}{a^3H}
\simeq \frac{\sqrt{2}\,a_{\text{nr}}}{a_{\text{eq}}^2 H_{\text{eq}}}\left[1-\frac{a_*}{a_{\text{nr}}}+\log\left(\frac{4 a_{\text{eq}}}{a_{\text{nr}}}\right)\right]\qquad\text{(relativistic)}\,.
\end{equation}
As noted in Sec.~\ref{sec:adiabatic}, the wave number $\lambda_{\text{FSH}}^{-1}$ represents the value of $k$ at which the matter power spectrum becomes suppressed compared to $\Lambda$CDM. This suppression can again be constrained by Ly-$\alpha$observations. We can directly compare the predicted free-streaming horizon in Eq.~\eqref{eq:lfshnr} (non-relativistic dark matter production) or Eq.~\eqref{eq:lfsh2} (relativistic dark matter production) with the maximal $\lambda_{\text{FSH}}$ imposed by WDM simulations~\cite{Garzilli:2019qki} (cf.\ Eq.~\eqref{eq:LFSH}). Requiring $\lambda_{\text{FSH}}<\lambda_{\text{FSH,max}}$ yields a bound on $a_*$ in the case of non-relativistic dark matter production and on $a_{\text{nr}}$ in the case of relativistic dark matter production,
\begin{align}\label{eq:fsconstraint}
a_{*}\,\log\left(\frac{0.0012}{a_*}\right) &< 3.5\times 10^{-6}\,\left(\frac{0.1}{v_*}\right)&\text{(non-relativistic production),}\\
a_{\text{nr}} &< 3.0\times 10^{-8} &\text{(relativistic production)}\,,
\label{eq:fsconstraint2}
\end{align}
where we neglected the dependence on $a_*$ in the relativistic case which affects the bound only at the percent level.

We find that in the case of non-relativistic dark matter production with $v_*<0.1$, the dominant constraint on $a_*$ comes from the requirement that observable scales are inside the horizon at the Dark Big Bang. Thus once we impose Eq.~\eqref{astmax}, the free-streaming constraint is automatically satisfied. In the opposite regime of highly relativistic dark matter production, free-streaming sets the leading constraint on the scale factor and we need to impose Eq.~\eqref{eq:fsconstraint2}.
In this case, we find that dark matter needs to become non-relativistic at
\begin{equation}
\label{eq:TNR}
t_{\text{nr}} < 0.3 \:\text{days}\quad \Longleftrightarrow \quad T_{\text{nr}} > 7.8\:\text{keV}
\qquad\qquad
\text{(relativistic production)}
\,.
\end{equation}
Here $T_{\text{nr}}$ denotes the (visible sector) temperature of the Universe at which the dark matter becomes non-relativistic. 

We can summarize the results of the previous and this section by saying that the Dark Big Bang must induce dark matter which is non-relativistic at a time $t= \mathcal{O}(\text{day})$ -- either by producing non-relativistic dark matter right away, or by producing relativistic dark matter which has redshifted to non-relativistic velocities by this time.

\section{A Model Realization of the Dark Big Bang}\label{sec:model}

In this section, we construct a model realization of the Dark Big Bang.  We write down a Lagrangian for the dark sector tunneling and dark matter fields.
Then we compute the tunneling rate for this Lagrangian.
We leave detailed models for the dark matter to the next section, Sec.~\ref{sec:dmproduction}.

\subsection{The Dark Sector}

The dark sector is comprised of the tunneling scalar $\phi$ and the dark matter particle $\chi$. The phenomenology described in this section hardly depends on whether $\chi$ is identified with a scalar, a fermion or a vector particle. For concreteness we take $\chi$ to be a real scalar in most parts of this work. 
Optionally, additional light /massless dark radiation degrees of freedom $\xi_i$ may be present.  The dark sector Lagrangian can thus be written as\footnote{We neglect a possible cubic term $\propto\phi \chi^2$ which is irrelevant for the following discussion.},
\begin{equation}\label{eq:LDS}
\mathcal{L}_{\text{DS}}= \frac{1}{2}\partial_\mu \phi\partial^\mu \phi + \frac{1}{2}\partial_\mu \chi\partial^\mu \chi  - V(\phi) - y\,\phi^2 \chi^2 - \frac{m_\chi^2}{2}\chi^2 -\kappa \chi^4 \left(+ \mathcal{L}_{\text{DR}}\right)\,,
\end{equation}
where we included only even powers of $\chi$ in order to ensure dark matter stability. This can be realized through a simple $Z_2$ symmetry under which $\chi$ is odd~\cite{Silveira:1985rk,Burgess:2000yq}. The Lagrangian part $\mathcal{L}_{\text{DR}}$ stands for terms involving the dark radiation field(s). We have put this term in brackets since we will consider cases with and without dark radiation (for the latter $\mathcal{L}_{\text{DR}}$ is absent).

The potential of the scalar $\phi$ is taken to be of generic quartic form,
\begin{equation}\label{eq:Vphi}
V(\phi)= \frac{m^2}{2}\phi^2-\mu\phi^3+\lambda \phi^4+\Delta V\,.
\end{equation}
In the following we assume $\mu>\sqrt{2\,\lambda}\,m$. For this choice the potential possesses a metastable minimum at $\phi=0$ with energy density $\Delta V$. We pick $\Delta V$ such that the energy density vanishes in the true minimum of the potential which is located at $\phi_{\text{min}}=\Delta\phi$. We thus have
\begin{align}
\Delta\phi &= \frac{3\mu+\sqrt{9\mu^2-16\lambda m^2}}{8\lambda}\,,\nonumber\\
\Delta V&= \frac{\left(3\mu+\sqrt{9\mu^2-16\lambda m^2}\right)^2\;\left(3\mu^2 + \mu \sqrt{9\mu^2-16\lambda m^2}-8\lambda m^2\right)}{2048\lambda^3}\,.
\end{align}
The mass of the scalar in the false vacuum is given by $m$, while the mass in the true vacuum is,
\begin{equation}
\label{eq:truevacmass}
m_\phi^2=V^{\prime\prime}(\Delta\phi) = \frac{9\mu^2 + 3\mu\sqrt{9\mu^2-16\lambda m^2}}{8\lambda} -2 m^2\,.
\end{equation}
The particle $\chi$ also receives a mass shift between the false and true vacuum due to the coupling term $y\,\phi^2 \chi^2$. The examples we will discuss in this work, however, feature $m_\chi^2\gg y(\Delta\phi)^2$ such that the dark matter mass shift can be ignored. Furthermore, we assume that the mass of the dark radiation (if present) generated in the true vacuum is negligible against the dark sector temperature such that it does not affect the cosmological evolution. 

\subsection{Tunneling Rate}

The tunneling rate from the false into the true vacuum reads~\cite{Coleman:1977py,Callan:1977pt} (cf.~Eq.~\eqref{eq:bounce1} and~\eqref{eq:bounce2}),
\begin{equation}
\Gamma\simeq m^4 \left(\frac{S}{2\pi}\right)^2\, e^{-S}\,,
\end{equation}
where the Euclidean action $S$ has to be determined by solving the differential equation of the bounce. An analytic solution exists in the thin-wall regime of vacuum tunnelling -- which is approached for a small energy difference $\Delta V \ll V_b$ between minima~\cite{Coleman:1977py}. However, while the thin-wall approximation is frequently used in the literature for its simplicity, it hardly ever applies to realistic tunneling phenomena. This is because in the thin-wall regime the lifetime of the false vacuum is typically so suppressed, that tunneling practically never occurs during the lifetime of the Universe.

Luckily, for the case of a quartic tunneling potential as considered in this work, an accurate numerical approximation of $S$ beyond the thin-wall regime has been derived~\cite{Adams:1993zs},
\begin{equation}\label{eq:thinw}
S\simeq \frac{\pi^2\mu^6}{24\lambda(\mu^2-2\lambda m^2)^3}\left[A_1\, \frac{4\lambda m^2}{\mu^2} +A_2 \,\left(\frac{4\lambda m^2}{\mu^2}\right)^2 + A_3 \,\left(\frac{4\lambda m^2}{\mu^2}\right)^3\right]\,,
\end{equation}
with $A_1=13.832$, $A_2=- 10.819$ and $A_3=2.0765$. The term in square brackets accounts for the deviation from the thin wall approximation and was obtained by a fit to the exact numerical bounce solution. If the Dark Big Bang occurs between BBN and matter-radiation equality $\Gamma^{-1/4} \simeq 1\text{s} - 10^4\text{yr}$ this corresponds to a bounce action in the range $S\simeq 190-250$. 

The tunneling potential also determines the radius $R_0$ of the true vacuum bubbles at nucleation~\cite{Coleman:1977py,Coleman:1980aw},
\begin{equation}\label{eq:nucradius}
R_0 \simeq \frac{m(\Delta \phi)^2}{2\,\Delta V}\,.
\end{equation}
Hence, $R_0$ is set by the typical energy scale of the tunneling potential. The above expression for $R_0$ strictly holds in the thin-wall approximation. Corrections to the thin-wall approximation change $R_0$ by an $\mathcal{O}(1)$-factor. We neglect this subtlety since a rough estimate of $R_0$ is sufficient for our purposes.

\section{Dark Matter Candidates from the Dark Big Bang}\label{sec:dmproduction}

In this section we will study dark matter production from the Dark Big Bang. While it is impossible to cover all properties and evolutions of the dark matter component, we will aim at reflecting a range of interesting cases. For this purpose we will study several example scenarios in which the dark matter abundance of the Universe is successfully reproduced. Other dark matter realizations in connection with a first-order phase transitions have been discussed, for instance, in~\cite{Watkins:1991zt,Chung:1998ua,Kolb:1998ki,Falkowski:2012fb,An:2022toi,Azatov:2021ifm,Shelton:2010ta,Petraki:2011mv,Krylov:2013qe,Baldes:2017rcu,Hall:2019rld,Huang:2017kzu,Hong:2020est,Kawana:2021tde,Kawana:2022lba,Witten:1984rs,Frieman:1990nh,Zhitnitsky:2002qa,Oaknin:2003uv,Lawson:2012zu,Atreya:2014sca,Bai:2018vik,Bai:2018dxf,Baker:2019ndr,Chway:2019kft}.

We note that particle production by a phase transition is a complicated process. In the first step, the energy of the false vacuum is converted to the kinetic energy of the bubble walls. Then, upon collision particles are either produced directly or via radiation of classical scalar waves which subsequently decay into particles~\cite{Watkins:1991zt}.

We will first turn to the case of light dark matter particles with mass $m_\chi\lesssim (\Delta V)^{1/4}$ and unsuppressed couplings to the tunneling field. The light particles are efficiently produced by the bubble collisions and typically reach a thermal state characterized by its temperature $T_{\text{DS}}$ soon after the phase transition. The dark temperature is given by $T_{\text{DS}}\sim (\Delta V)^{1/4}$ up to $\mathcal{O}(1)$-factors. In this scenario, the details of the phase transition do not enter the calculation of the final dark matter density since dark matter annihilation and scattering reactions quickly establish a thermal spectrum and wash out any characteristics of the state emerging from the bubble collisions.

In Sec.~\ref{sec:darkzilla} we will then turn to the opposite case of heavy dark matter ($m_\chi\gg (\Delta V)^{1/4}$) production. The heavy dark matter particles emerging from the bubble collisions feature a non-thermal spectrum. Due to the small number density of the dark matter particles, number changing reactions are typically inefficient and the dark matter number remains frozen after the Dark Big Bang. This case is conceptually more involved since the final dark matter density is set by the microscopic details of particle formation by the bubble collisions.

\subsection{Light Dark Matter}\label{sec:lightdm}

In this section we consider dark matter particles $\chi$ with mass below the energy scale of the Dark Big Bang, $m_\chi \lesssim (\Delta V)^{1/4}$. We will focus on the simplest case, in which $\chi$ enters a thermal equilibrium state shortly after the Dark Big Bang. Since the evolution of the dark matter density depends on whether or not dark radiation degrees of freedom (coupled to the dark matter) also exist in the dark sector, we will discuss both possibilities separately. We will see that in both cases the relic density is set by a thermal freeze-out. However, differences emerge since $\chi$ runs through a period of cannibalism~\cite{Carlson:1992fn} in the absence of a dark radiation bath (or if the dark radiation is only extremely weakly coupled to dark matter), while $\chi$ undergoes standard pair-annihilations in the presence of a dark radiation bath -- analogous to WIMPs~\cite{Dicus:1977nn,Lee:1977ua,Hut:1977zn}.

Previous papers have studied versions of cannibal dark matter~\cite{Carlson:1992fn,Pappadopulo:2016pkp} and dark sector WIMP dark matter~\cite{Feng:2008mu,Berlin:2016gtr} under the assumption of the existence of a thermal bath in the dark sector (for instance from asymmetric reheating after inflation~\cite{Hodges:1993yb,Berezhiani:1995am,Adshead:2016xxj}). In our work, on the other hand, the Dark Big Bang sets the initial conditions for the dark matter. As we will show, the specifics of the tunneling field set the phase transition parameters, which then allow explicit calculation of the initial dark matter density and temperature. Further, as an example, we will present complete scenarios with specific benchmark points for the parameters of the tunneling field and resulting dark matter properties (see Tab.~\ref{tab:cannibalbp} and~\ref{tab:rhodarkwimp} below). Our study of the production and evolution of the dark cannibals and dark WIMPs in the context of a Dark Big Bang follows.

\subsubsection{Dark Cannibals}\label{sec:cannibal}

We first consider a minimal scenario in which the dark sector only contains the tunneling field $\phi$ and the light dark matter scalar $\chi$. We take $m_{\chi} < m_{\phi}$. The Lagrangian is given by Eq.~\eqref{eq:LDS} without the $\mathcal{L}_{\text{DR}}$ term (since we do not introduce any dark radiation fields for now). We dub the dark matter particles ``dark cannibals'' in this scenario since they cannibalize themselves for some period of the cosmological evolution as we will describe below.

We argued in Sec.~\ref{sec:darkcosmology} that $\phi$ remains trapped in the false vacuum during the early radiation-dominated epoch. The dark sector remains cold until the moment when the Dark Big Bang phase transition generates a hot plasma of $\chi$- and $\phi$-particles. Since particle number changing reactions are very efficient shortly after the phase transition (unless for highly suppressed couplings $\lambda,\,\kappa\lesssim 10^{-10}$) the details of particle production by the bubble collisions can be ignored. Instead, a thermal equilibrium spectrum is quickly established. The dark reheating temperature $T_{\text{DS},*}$ (= the dark plasma temperature right after the Dark Big Bang) can be estimated as (cf.~Eq.~\eqref{eq:darkreheating}),
\begin{equation}\label{eq:TDSstar}
T_{\text{DS},*} \simeq \left(\frac{30}{\pi^2\,g_{\text{DS}}(T_{\text{DS},*})}\Delta V\right)^{1/4}\,,
\end{equation}
where $g_{\text{DS}}$ counts the number of relativistic degrees of freedom in the dark sector which includes $\chi$ and, possibly, $\phi$. Because typically $m_\phi \sim (\Delta V)^{1/4} \sim T_{\text{DS},*}$, $\phi$ is often semi-relativistic after the Dark Big Bang -- thus contributing a fractional degree to $g_{\text{DS}}$.

We emphasize that $T_{\text{DS},*}$ is generically different from the visible sector temperature at the Dark Big Bang $T_*$. While $T_{\text{DS},*}$ is set by the energy scale of the phase transition, $T_*$ depends on the time of the phase transition and, hence, the tunneling rate (cf. Eq.~\eqref{eq:tstar} and~\eqref{eq:Tstar}). Since dark and visible sector are decoupled, the two temperatures evolve independently. The entropies of visible and dark sector are separately conserved. Therefore, it is convenient to introduce the ratio of dark-to-visible entropy $\xi$
\begin{equation}\label{eq:xi}
\xi=\frac{s_{\text{DS}}}{s} = \frac{g_{\text{DS}}(T_{\text{DS},*})\,T_{\text{DS},*}^3}{g_{\text{eff}}(T_*)\,T_*^3}\,,
\end{equation}
with $T_*$ and $T_{\text{DS},*}$ given by Eq.~\eqref{eq:Tstar} and~\eqref{eq:TDSstar} respectively. The entropy ratio $\xi$ is conserved during the post-Dark-Big-Bang evolution of the Universe.

\begin{figure}[htp]
\begin{center}
\includegraphics[width=0.25\textwidth]{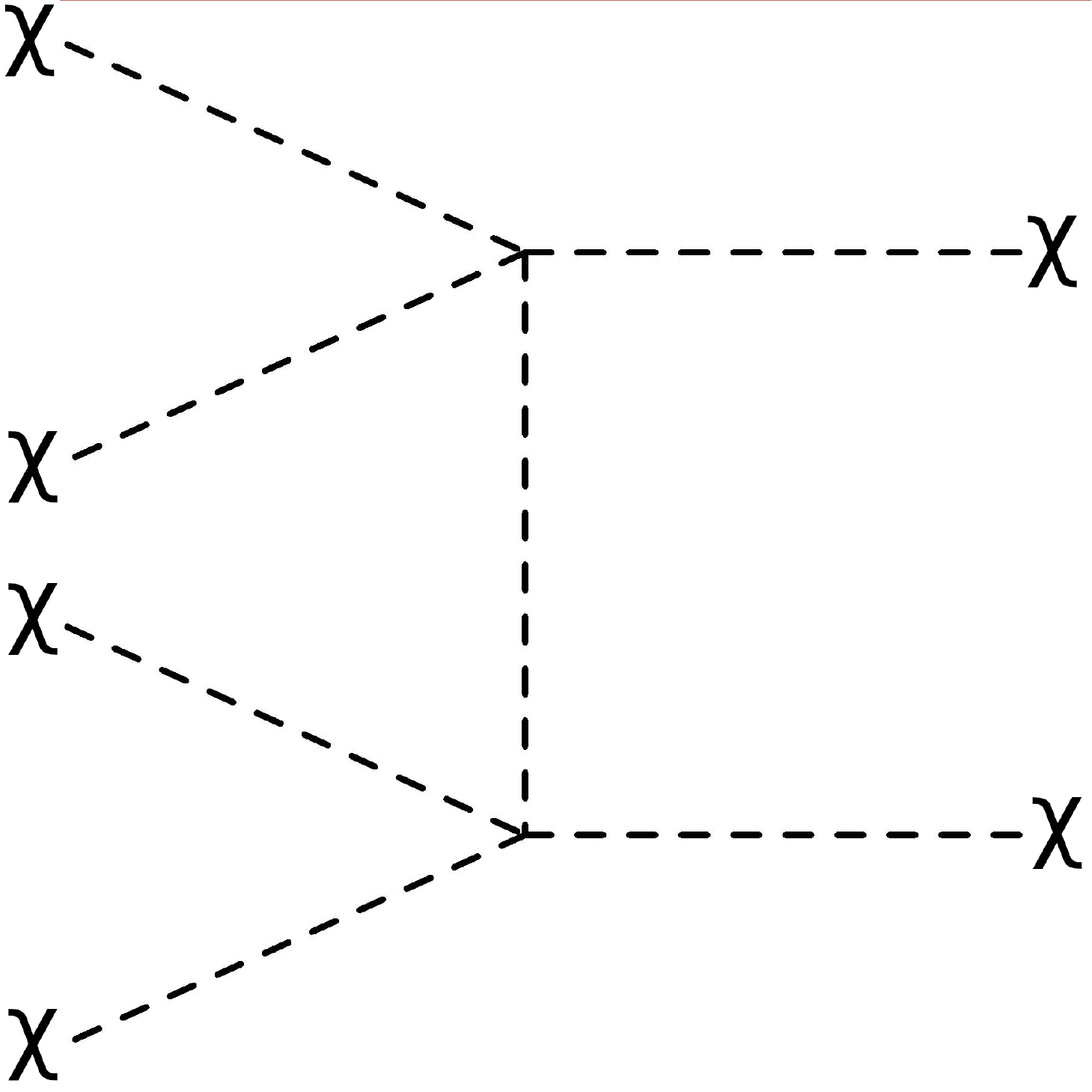}
\end{center}
\vspace{-4mm}
\caption{Feynman Diagram for $\chi\chi\chi\chi\leftrightarrow \chi\chi$ reactions.}
\label{fig:feynman}
\end{figure}

Shortly after the Dark Big Bang interconversion reaction $\phi\phi \leftrightarrow \chi \chi$ are typically still active and both species contribute significantly to the energy density. However, once $T_{\text{DS}}\ll m_\phi$ the production of $\phi$ gets strongly Boltzmann suppressed. The existing $\phi$-particles quickly annihilate/ decay away leaving all the dark sector energy density in $\chi$. At this stage -- in the absence of any other light degrees of freedom -- $\chi$ can no longer undergo pair-annihilations. However -- given a quartic coupling $\kappa \gtrsim 10^{-5}$ -- $\chi\chi\chi\chi\leftrightarrow \chi\chi$ processes as shown in Fig.~\ref{fig:feynman} still keep $\chi$ in thermal equilibrium for some time. Related scenarios -- albeit without a Dark Big Bang origin of $\chi$ -- have previously been discussed in~\cite{Bernal:2015xba,Heikinheimo:2016yds,Bernal:2017kxu,Arcadi:2019oxh}.

\noindent
The number density of $\chi$ can be estimated as~\cite{Gondolo:1990dk},
\begin{equation}\label{eq:neq}
n_{\chi}(T_{\text{DS}})\simeq \frac{m_\chi^2 T_{\text{DS}}}{2\pi^2}\,e^{\mu/T_{\text{DS}}}\, K_2\left(\frac{m_\chi}{T_{\text{DS}}}\right)\,,
\end{equation} 
where $K_2$ denotes the modified Bessel function of the second kind. In thermal equilibrium the chemical potential $\mu_\chi$ vanishes, i.e.\ the equilibrium density is given by $n_{\chi,\text{eq}}(T_{\text{DS}}) =n_{\chi}(T_{\text{DS}},\mu_\chi=0)$.
The expression~\eqref{eq:neq} was derived by approximating the Bose-Einstein phase space distribution by a Maxwell-Boltzmann distribution. The corresponding error on $n_{\chi,\text{eq}}$ is at most $20\% $ (and quickly approaches zero for $T_{\text{DS}}< m_\chi$).
Once the $\phi$-particles have decayed away the dark sector entropy is entirely carried by $\chi$. The entropy density can be approximated as\footnote{In order to arrive at this expression we have again approximated the Bose-Einstein phase space distribution by a Maxwell-Boltzmann distribution. The corresponding error on $s$ is $<10\%$ in the relativistic regime and quickly approaches zero in the non-relativistic regime.},
\begin{equation}\label{eq:sdark}
s_{\text{DS}} \simeq \left(\frac{m_\chi}{T_{\text{DS}}}\frac{K_3\left(\frac{m_\chi}{T_{\text{DS}}}\right)}{K_2\left(\frac{m_\chi}{T_{\text{DS}}}\right)}-\frac{\mu_\chi}{T_{\text{DS}}} \right) n_\chi\,.
\end{equation}

A peculiarity occurs if the number changing reactions are still active once $\chi$-particles become non-relativistic. In the non-relativistic regime during thermal equilibrium ($\mu_\chi\simeq 0$) the dark entropy density approaches,
\begin{equation}
s_{\text{DS}} \xrightarrow{T_{\text{DS}}\ll m_\chi}\;\; \frac{m_\chi^{5/2} \, T^{1/2}_{\text{DS}}}{(2\pi)^{3/2}} e^{-m_\chi/T_{\text{DS}}}\,.
\end{equation}
Since the total dark entropy $\propto s_{\text{DS}} a^3$ needs to be conserved, $T_{\text{DS}}$ decreases only logarithmically with the scale factor of the Universe, $T_{\text{DS}}\propto 1/\log(a)$. This is in contrast to the visible sector, whose entropy is always dominated by relativistic degrees of freedom,
\begin{equation}\label{eq:svis}
s = \frac{2\pi^2}{45}\,g_{\text{eff}}(T)\,T^3\,.
\end{equation}
Consequently, the visible sector temperature decreases as $T\propto 1/a$. This implies that $T_{\text{DS}}/T$ increases once $\chi$ enters the non-relativistic regime (assuming it is still in thermal equilibrium at that time). Depending on the specific parameter choice the dark sector may even become hotter than the visible sector for some time. This can be understood by the $\chi\chi\chi\chi\rightarrow \chi\chi$ reactions of Fig.~\ref{fig:feynman} which convert non-relativistic $\chi$-particles into fewer relativistic $\chi$-particles. The excess kinetic energy is then quickly distributed among the bath of $\chi$-particles to keep it warm. In~\cite{Carlson:1992fn} this phenomenon was dubbed cannibalism since the dark matter particles cannibalize their rest mass for staying warm. We shall thus refer to the $\chi$-particles as ``dark cannibals'' in this section. Let us note, however, that the cannibalism stage does not imply a realization of warm (or even hot) dark matter since it occurs when the dark cannibals are already non-relativistic.

Once the rate for $\chi\chi\chi\chi\leftrightarrow \chi\chi$ reactions drops below the Hubble rate of expansion, $\chi$ freezes out of thermal equilibrium and the total number of dark cannibals in the Universe remains fixed. In order to derive the evolution of the number density $n_\chi$ we need to solve the Boltzmann equation~\cite{Arcadi:2019oxh}
\begin{equation}\label{eq:boltzmann1}
\frac{dn_\chi}{dt} + 3 H n_\chi = 2 \left(\Gamma_{\chi\chi\rightarrow \chi\chi\chi\chi} -\Gamma_{\chi\chi\chi\chi\rightarrow \chi\chi}   \right)\,.
\end{equation}
In the non-relativistic regime, the rate of $4\rightarrow 2$ and $2\rightarrow 4$ processes is given as
\begin{equation}\label{eq:GammaSSSS}
\Gamma_{\chi\chi\chi\chi\rightarrow \chi\chi} = e^{2\mu/T_{\text{DS}}}\,\Gamma_{\chi\chi\rightarrow \chi\chi\chi\chi} =\langle \sigma_{\chi\chi\chi\chi\rightarrow \chi\chi} \,v^3\rangle\, n_\chi^4\,,
\end{equation}
where the thermally averaged annihilation cross section for the model under consideration (cf.~\eqref{eq:LDS}) reads\footnote{We are assuming that the cross section is dominated by processes mediated by the quartic coupling $\mathcal{L}_{\text{DS}}\supset \kappa \chi^4$ as shown in Fig.~\ref{fig:feynman}. An additional contribution to the cross section emerges from processes with $\phi$ in the intermediate state. However, around the time when $\chi$ freezes out from equilibrium, $\phi$ can be integrated out in the Lagrangian~\eqref{eq:LDS}. In the resulting effective Lagrangian, there then appears an additional quartic term $\sim y^2\,(\langle\phi\rangle/m_\phi)^2\, \chi^4$ which we neglected. Even if this additional term were non-negligible it can be absorbed by a redefinition of $\kappa$. Hence, Eq.~\eqref{eq:sigmav3} would still hold if one replaces $\kappa$ by the redefined $\kappa$.}~\cite{Arcadi:2019oxh}
\begin{equation}\label{eq:sigmav3}
\langle \sigma_{\chi\chi\chi\chi\rightarrow \chi\chi} \,v^3\rangle 
=\frac{27\sqrt{3}\kappa^4}{\pi m_\chi^8}\,.
\end{equation}
Notice the factor $v^3$ in the definition of the thermally-averaged cross section which occurs due to the four particles in the initial state. This is in contrast to 2-particle annihilation processes (occurring e.g.\ in standard WIMP scenarios) for which the thermally averaged cross section is defined with a single power of the velocity $v$. Furthermore, we point out that $\langle \sigma_{\chi\chi\chi\chi\rightarrow \chi\chi} \,v^3\rangle$ does not carry a temperature-dependence. This is because $\sigma_{\chi\chi\chi\chi\rightarrow \chi\chi} \,v^3$ turns out to be velocity-independent.

It is convenient to define the dark cannibal abundance as 
\begin{equation}\label{eq:Y}
Y_\chi = \frac{n_\chi}{s}\,,
\end{equation}
where $s$ again stands for the visible sector entropy. Entropy conservation implies that the quantity $Y_\chi$ is conserved after the freeze-out of $\chi$.\footnote{Notice that it is just a matter of convention that we used $s$ and not the dark sector entropy $s_{\text{DS}}$ in the definition of $Y_\chi$.} We can then rewrite the Boltzmann equation in terms of $Y_\chi$, and use Eq.~\eqref{eq:GammaSSSS} to express the reaction rates in terms of the annihilation cross section, 
\begin{equation}\label{eq:boltzmann2}
\frac{dY_\chi}{dT} = \frac{2\,\langle \sigma_{\chi\chi\chi\chi\rightarrow \chi\chi}\, v^3 \rangle\, s^3}{T\,H}
\,\left(1+\frac{T}{3g}\frac{dg}{dT}\right)\,(Y_\chi^4- Y_\chi^2\,Y_{\chi,\text{eq}}^2)\,,
\end{equation}
where we also converted the time-derivative into a temperature-derivative by using Eq.~\eqref{eq:svis} and $d(s\,a^3)/dt=0$. In order to solve the above equation we first need to express the dark sector temperature and the chemical potential in terms of the visible sector temperature. Using Eq.~\eqref{eq:xi},~\eqref{eq:neq},~\eqref{eq:sdark},~\eqref{eq:svis} and~\eqref{eq:Y}, we obtain
\begin{equation}\label{eq:TDSmu}
T_{\text{DS}}= \frac{m_\chi}{\mathcal{F}^{-1}\left(\frac{4\pi^4 g_{\text{eff}}(T) T^3 Y_\chi}{45 \, m_\chi^3}\:e^{\xi/Y_\chi}\right)}\,,
\qquad
\frac{\mu_\chi}{T_{\text{DS}}} = \frac{m_\chi}{T_{\text{DS}}}\frac{K_3\left(\frac{m_\chi}{T_{\text{DS}}}\right)}{K_2\left(\frac{m_\chi}{T_{\text{DS}}}\right)} - \frac{\xi}{Y_\chi}\,,
\end{equation}
where $\mathcal{F}^{-1}$ is the inverse function of
\begin{equation}\label{eq:function}
\mathcal{F}(x)= \frac{K_2(x)}{x}\exp\left(\frac{x\,K_3(x)}{K_2(x)}\right)\,.
\end{equation}
While there is no analytic expression for $\mathcal{F}^{-1}$, one can easily obtain $\mathcal{F}^{-1}$ by inverting Eq.~\eqref{eq:function} numerically. From $Y_{\chi,\text{eq}}=e^{-\mu_s/T_{\text{DS}}}Y_\chi$ it follows that
\begin{equation}\label{eq:YSeq}
Y_{\chi,\text{eq}}= \exp\left(-\frac{m_\chi}{T_{\text{DS}}}\frac{K_3\left(\frac{m_\chi}{T_{\text{DS}}}\right)}{K_2\left(\frac{m_\chi}{T_{\text{DS}}}\right)} + \frac{\xi}{Y_\chi} \right)\,Y_\chi\,.
\end{equation}
Now we plug $T_{\text{DS}}$ from Eq.~\eqref{eq:TDSmu} into the above expression and insert the resulting $Y_{\chi,\text{eq}}$ into~\eqref{eq:boltzmann2}. In this way we obtain the Boltzmann equation for $Y_\chi$ in terms of a single variable $T$. The evolution of $Y_\chi$ can then be obtained by solving the Boltzmann equation numerically.

A further simplification occurs for $T_{\text{DS}}\ll m_\chi$. In this limit one can approximate the Bessel functions $K_{2,3}(x) \rightarrow \sqrt{\pi/(2x)}\,e^{-x}$ and obtain the following analytic expression for the equilibrium abundance (see~\cite{Arcadi:2019oxh}),
\begin{equation}\label{eq:YSeqap}
Y_{\chi,\text{eq}}\xrightarrow{T_{\text{DS}}\ll m_\chi} \exp\left(\frac{\xi}{Y_\chi}-1 - \frac{m_\chi^2}{2\pi T^2}\left(\frac{45}{2\pi^2 g_{\text{eff}}(T) Y_\chi}\:e^{1-\xi/Y_\chi}\right)^{2/3}  \right)\,Y_\chi\,.
\end{equation}
However, in this paper we refrain from using Eq.~\eqref{eq:YSeqap} in the Boltzmann equation.
While the final relic abundance can be accurately predicted using Eq.~\eqref{eq:YSeqap} in the case of a non-relativistic freeze-out, a significant error arises for a semi-relativistic freeze-out for which $1\lesssim m_\chi/T_{\text{DS},\text{FO}}\lesssim 3$. Here, $T_{\text{DS},\text{FO}}$ denotes the dark sector freeze-out temperature which can be defined by the condition $Y_{\chi,\text{eq}}(T_{\text{DS},\text{FO}}) = Y_{\chi,\infty}$.
Hence in this paper we do not use the  non-relativistic approximation of Eq.~\eqref{eq:YSeqap}.

Instead, we build on the previous literature by plugging the full expression Eq.~\eqref{eq:YSeq} into the Boltzmann equation to obtain an accurate prediction for $Y_{\chi,\infty}$ in both the non-relativistic as well as in the semi-relativistic freeze-out regime (this has been shown for an analogous case of WIMP dark matter in~\cite{Drees:2009bi}). 

Once we have derived $Y_{\chi,\infty}$, the dark cannibal relic density can be calculated,
\begin{equation}\label{eq:Omegachi}
\Omega_{\chi}h^2= \frac{m_\chi \; s(T_0) \;Y_{\chi,\infty}}{3 \,(H_0/h)^2\, M_P^2}
= 275 \: Y_{\chi,\infty} \, \left(\frac{m_\chi}{\text{keV}}  \right)\;.
\end{equation}

\begin{figure}[htp]
\begin{center}
\includegraphics[height=6.0cm]{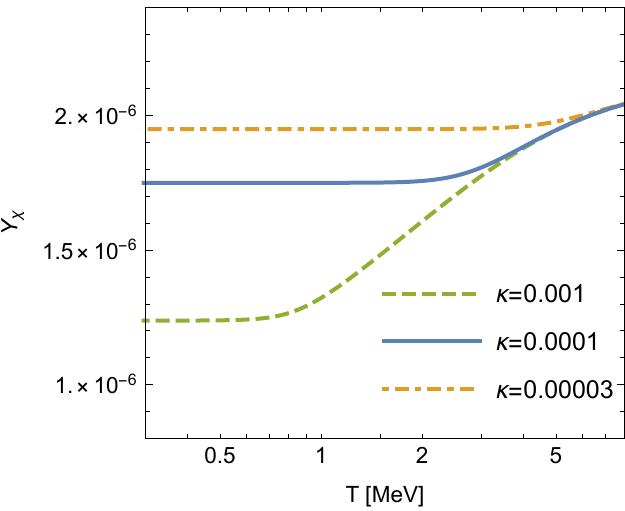}\hspace{6mm}
\includegraphics[height=6.0cm]{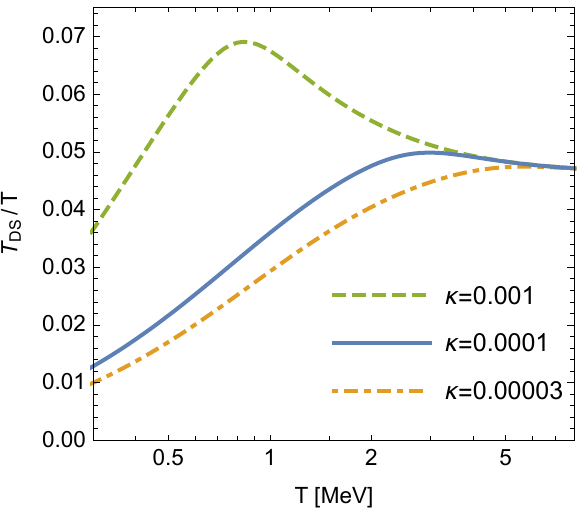}
\end{center}
\caption{Evolution of the dark cannibal abundance $Y_\chi$ (left panel) and the dark sector temperature $T_{\text{DS}}$ (right panel) after the Dark Big Bang as a function of the visible sector temperature $T$. The blue line in both panels corresponds to the parameter choice of Tab.~\ref{tab:cannibalbp}. The green dashed and orange dot-dashed lines refer to a larger and smaller choice of the dark cannibal self-coupling $\kappa$ as indicated in the plot legend (while the remaining input parameters were again taken from Tab.~\ref{tab:cannibalbp}). As can be seen, for larger $\kappa$ the dark cannibals can maintain thermal equilibrium longer and deplete their abundance more efficiently. Therefore, the dark cannibal relic abundance $Y_{\chi,\infty}$ scales inversely with $\kappa$.}
\label{fig:YTd}
\end{figure}

We now consider three benchmark cases with a dark cannibal mass $m_\chi=250\:\text{keV}$ and self-coupling $\kappa=0.001,\,0.0001,\,0.00003$. The remaining input parameters are taken from Tab.~\ref{tab:cannibalbp}. All three cases feature a Dark Big Bang at a visible sector temperature of $T_*=8\:\text{MeV}$ which heats the dark sector to a temperature $T_{\text{DS},*}=0.4\:\text{MeV}$. The subsequent temperature-evolution of the dark cannibal abundance $Y_\chi$ obtained by solving the Boltzmann equation~\eqref{eq:boltzmann2} is shown in Fig.~\ref{fig:YTd} (left panel). It can be seen that $Y_\chi$ decreases as the Universe cools down until $\chi\chi\chi\chi\rightarrow \chi\chi$ reactions freeze out of equilibrium. After the freeze-out $Y_\chi$ quickly becomes constant and reaches the final relic abundance $Y_{\chi,\infty}$. As expected, comparison of the three benchmark cases reveals that the freeze-out occurs the later the larger $\kappa$ (since a large self-coupling is needed to keep $\chi$ in thermal equilibrium). Consequently, the dark cannibal relic density scales inversely with $\kappa$.

In the right panel of Fig.~\ref{fig:YTd} the evolution of the temperature ratio $T_{\text{DS}}/T$ is shown. Shortly after the Dark Big Bang (on the right side of the figure) when $\chi$ is still relativistic the ratio $T_{\text{DS}}/T$ remains approximately constant. But once $\chi$ enters the non-relativistic regime (at $T\sim 5\:\text{MeV}$ in the figure) the dark sector temperature increases relative to the visible sector temperature. This is the described epoch in which $\chi$ cannibalizes its rest mass in order to keep the dark temperature nearly constant. The increase of $T_{\text{DS}}/T$ is clearly visible for the benchmark case with $\kappa= 0.001$, visible but less pronounced for $\kappa= 0.0001$ and not visible at all for $\kappa= 0.00003$. This is because the cannibalistic epoch only occurs if $\chi$ is still in thermal equilibrium when it becomes non-relativistic. This condition is satisfied for the two cases with larger $\kappa$, while for $\kappa=0.00003$ the freeze-out occurs right at the transition between the relativistic and the non-relativistic regime ($T_{\text{DS},\text{FO}}\simeq m_\chi$). After the freeze-out $T_{\text{DS}}/T$ drops quickly in all three benchmarks. This is because the temperature of a decoupled non-relativistic species decreases as $T_{\text{DS}}\propto 1/a^2$ with the scale factor while $T\propto 1/a$ for the visible sector (which contains relativistic degrees of freedom).

\begin{table}[htp]
\begin{center}
\begin{tabular}{|ll|ll|}
\hline
&&&\\[-4mm]
\multicolumn{2}{|c|}{Input Parameters}& \multicolumn{2}{c|}{Phase Transition} \\
 \hline
&&&\\[-4mm] 
$m_\chi$~[keV] & $250$ & $t_*$~[s] & $0.012$ \\[1mm]
$m$~[keV] & $512$ & $T_*$~[MeV] & $8.0$\\[1mm]
$\mu$~[keV] & $802$ & $T_{\text{DS},*}$~[MeV] & $0.37$ \\[1mm]
$\kappa$ & $0.0001$ & $\alpha$ & $3.9\times 10^{-7}$\\
\cline{3-4}
&&&\\[-4mm]
$\lambda$ & $1$ & \multicolumn{2}{c|}{Dark Matter} \\
\hline
&&&\\[-4mm]
\multicolumn{2}{|c|}{Derived Parameters} & $m_\chi/T_{\text{DS},\text{FO}}$ & $1.3$ \\
\cline{1-2}
&&&\\[-4mm]
$m_\phi$~[keV] & $761$ & $\Omega_\chi h^2$ & $0.120$ \\[1mm]
$(\Delta V)^{1/4}$~[keV] & $273$ & $\sigma/m_\chi$~[$\text{cm}^2/\text{g}$] & $0.2$\\
\hline
\end{tabular}
\end{center}
\vspace{-0.4cm}
\caption{Parameter Example for a Dark Big Bang inducing dark cannibal dark matter. Here $m_\chi$ is the dark matter mass, parameters $m, \mu, \lambda$ of  the tunneling potential are as defined in Eq.~\eqref{eq:Vphi}, the mass of the tunneling field in the true vacuum $m_\phi$ is given in Eq.~\eqref{eq:truevacmass}, and $\kappa$ is the quartic self-coupling of the cannibal dark matter (see Eq.~\eqref{eq:LDS}).}
\label{tab:cannibalbp}
\end{table}

In the following we focus on the benchmark case with $\kappa=0.0001$ for which we also provide the phase transition parameters and dark matter properties in Tab.~\ref{tab:cannibalbp}. For this benchmark point the dark cannibal relic density $\Omega_\chi h^2$ exactly matches the observed dark matter density of $\Omega_{\text{DM}} h^2=0.120\pm 0.001$~\cite{Planck:2018vyg}. This shows that dark cannibals arising from a Dark Big Bang constitute an excellent candidate for the dark matter of the Universe. Another interesting observation concerns the cross section for dark matter self-scattering $\chi\chi\rightarrow \chi\chi$ which is given as
\begin{equation}
\sigma = \frac{9\kappa^2}{2\pi\,m_\chi^2}\,.
\end{equation}
For the benchmark point of Tab.~\ref{tab:cannibalbp} we obtain $\sigma/m_\chi =0.2 \text{cm}^2/g$. But more generally -- 
if the dark cannibals are light -- they can play the role of self-interacting dark matter~\cite{Spergel:1999mh}. For instance, for $m_\chi\lesssim \text{MeV}$ we find 
\begin{equation}
\sigma/m_\chi \gtrsim 10^{-3} \text{cm}^2/g\,,
\end{equation}
if we impose the dark cannibal scenario (i.e.\ if we impose that $\chi\chi\chi\chi\rightarrow \chi\chi$ reactions are active after the Dark Big Bang).

The most sensitive probes of dark matter self-interactions currently arise from the gravitational lensing of density profiles in galaxy clusters. Observation based on this technique place the cross section upper limit in the range $(\sigma/m_\chi)_\text{max}=0.2-0.4\:\text{cm}^2/g$ and suggest a very mild preference for a non-zero cross section of $\sigma/m_\chi\simeq 0.1-0.2\: \text{cm}^2/g$~\cite{Sagunski:2020spe,Andrade:2020lqq} albeit with some systematic uncertainties.\footnote{See also~\cite{Adhikari:2022sbh} for a recent review on astrophysical probes of dark matter self-interactions.} Excitingly, future searches for dark matter self-interactions will thus test a significant part of the dark cannibal parameter space.\footnote{See~\cite{Wang:2022akn} for another realization of self-interacting dark matter from a first-order phase transition.} If the cross section is as large as in the benchmark point a discovery of dark matter self-interactions with the next generation of observations and analysis tools could be just around the corner.

\subsubsection{Dark WIMPs}

In this section we again consider a dark sector including the tunneling field $\phi$ and the dark matter field $\chi$ with mass $m_\chi\lesssim (\Delta V)^{1/4}$. But in addition, we introduce massless (or very light) dark radiation $\xi$. In the presence of dark radiation  $\xi$, the evolution of dark matter particles $\chi$ will resemble that of an ordinary thermal WIMP, which is why we call $\chi$ a dark WIMP in this scenario.
The particle nature of $\xi$ is not particularly important.\footnote{From a theoretical point of view it would be more natural to consider light/massless fermionic dark radiation since fermion masses are protected by chiral symmetry in contrast to scalar masses. However, the phenomenology we describe in this section is insensitive to the particle nature of $\xi$ and, therefore, we made the simplest choice of a real scalar $\xi$.} Therefore, we make the simplest choice of a single real scalar degree of freedom $\xi$. In order to keep the discussion minimal, we furthermore impose a $Z_2$ symmetry on $\xi$ (in addition to the $Z_2$-symmetry on $\chi$) such that the Lagrangian only contains even powers of $\xi$, $\chi$. The full Lagrangian is again given by Eq.~\eqref{eq:LDS}, this time including the $\mathcal{L}_{\text{DR}}$ term with,
\begin{equation}\label{eq:LDR}
 \mathcal{L}_{\text{DR}} \supset y^\prime \chi^2 \xi^2\,.
\end{equation}
An additional dark radiation self-coupling and an interaction term involving the tunneling field are typically present, but do not affect the following discussion (as long as we can treat $\xi$ as effectively massless).

Let us now turn to the cosmological evolution. As described in Sec.~\ref{sec:cannibal} the Dark Big Bang heats the initially cold dark sector to the temperature $T_{\text{DS},*}$ determined by Eq.~\eqref{eq:TDSstar}. The resulting hot dark sector plasma contains $\chi$ and $\xi$ in thermal equilibrium. In the presence of dark radiation  $\xi$, the dark matter number changing reactions are dominantly pair-annihilations $\chi\chi \longleftrightarrow \xi\xi$, while $\chi\chi\chi\chi\longleftrightarrow \chi\chi$ processes play no role (unless $y^\prime\ll \kappa$). Once the pair-annihilation rate drops below the Hubble rate of expansion, $\chi$ freezes out and the total number of dark WIMPs in the Universe remains fixed. Related scenarios of dark matter undergoing a freeze-out in a decoupled dark sector have been discussed in~\cite{Feng:2008mu,Ackerman:2008kmp,Sigurdson:2009uz,Das:2010ts,Boddy:2014yra,Berlin:2016gtr}.

The entropies of dark and visible sector are separately conserved. However, in contrast to the dark cannibal scenario, the dark sector contains relativistic degrees of freedom during the entire post-Dark-Big-Bang evolution which dominate $s_{\text{DS}}$. As a consequence the ratio of dark-to-visible temperature remains fixed up to changes in the number of degrees of freedom,
\begin{equation}\label{eq:TdT}
 \frac{T_{\text{DS}}}{T} = \left(\frac{g_{\text{eff}}(T)}{g_{\text{eff}}(T_*)}\right)^{1/3}\left(\frac{g_{\text{DS}}(T_{\text{DS},*})}{g_{\text{DS}}(T_{\text{DS}})}\right)^{1/3}\frac{T_{\text{DS},*}}{T_*}\,,
\end{equation}
where $T_*$ and $T_{\text{DS},*}$ are obtained from Eq.~\eqref{eq:Tstar} and~\eqref{eq:TDSstar} respectively. 

The Boltzmann equation is conveniently expressed in terms of the dark WIMP abundance $Y_\chi=n_\chi/s$. One obtains,
\begin{equation}\label{eq:boltzmannwimp}
\frac{dY_\chi}{dT} = \frac{\langle \sigma_{\chi\chi\rightarrow \xi\xi}\, v \rangle\, s}{T\,H}
\,\left(1+\frac{T}{3g_{\text{eff}}}\frac{dg_{\text{eff}}}{dT}\right)\,(Y_\chi^2- Y_{\chi,\text{eq}}^2)\,,
\end{equation}
where $\langle \sigma_{\chi\chi\rightarrow \xi\xi}\, v \rangle$ denotes the thermally averaged dark WIMP annihilation cross section. For the interaction term in Eq.~\eqref{eq:LDR} the cross section is~\cite{McDonald:1993ex,Burgess:2000yq}
\begin{equation}
  \langle\sigma_{\chi\chi\rightarrow \xi\xi}\, v \rangle = \frac{(y^\prime)^2}{4\pi m_\chi^2}\,,
\end{equation}
in the non-relativistic regime. 

While Eq.~\eqref{eq:boltzmannwimp} looks identical to the Boltzmann equation of a standard thermal WIMP, an important difference is that the equilibrium number density of $\chi$ is set by the dark rather than the visible sector temperature. Specifically, we have
\begin{equation}\label{eq:YeqDW}
Y_{\chi,\text{eq}} =  \frac{m_\chi^2 T_{\text{DS}}}{2\pi^2\,s}\, K_2\left(\frac{m_\chi}{T_{\text{DS}}}\right)\,,
\end{equation}
where we again approximated the Bose-Einstein phase space distribution by a Maxwell-Boltzmann distribution. After plugging Eq.~\eqref{eq:YeqDW} into Eq.~\eqref{eq:boltzmannwimp} and eliminating $T_{\text{DS}}$ via Eq.~\eqref{eq:TdT} we obtain the Boltzmann equation for $Y_\chi$ in terms of a single variable $T$. This equation can be solved numerically. In the non-relativistic regime the dark WIMP energy density is given by $\rho_\chi = m_\chi Y_\chi s$. The final relic density $\Omega_{\chi}h^2$ is obtained via Eq.~\eqref{eq:Omegachi}.

\begin{figure}[htp]
\begin{center}
\includegraphics[height=7.4cm]{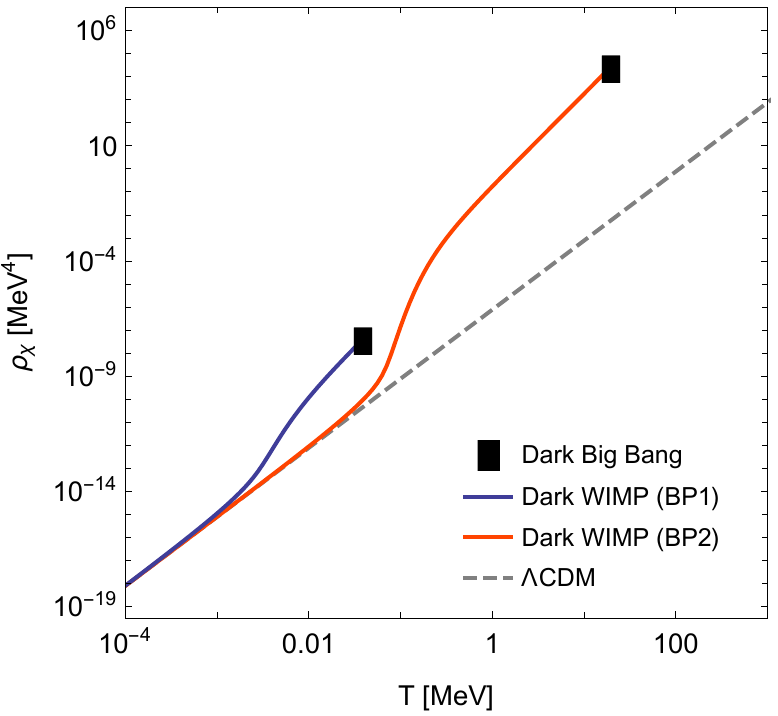}
\end{center}
\caption{Evolution of the dark WIMP energy density after the Dark Big Bang as a function of the visible sector temperature $T$. The two benchmark cases of Tab.~\ref{tab:rhodarkwimp} are depicted. For comparison, the dark matter energy density in $\Lambda$CDM is also shown.}
\label{fig:rhodarkwimp}
\end{figure}

\begin{table}[t]
\begin{center}
\begin{tabular}{|l|c|c|}
\hline
&&\\[-4mm]
Benchmark & BP1 & BP2 \\
 \hline
\multicolumn{3}{|c|}{$\,$}\\[-4mm] 
\multicolumn{3}{|c|}{Input Parameters} \\
 \hline
&&\\[-4mm] 
$m_\chi$~[keV] & $10$ & $500$\\[1mm]
$m$~[keV] & $25.5$ & $23455$\\[1mm]
$\mu$~[keV] & $39.6$ & $36683$\\[1mm]
$y^\prime\times 10^7$ & $0.011$ & $0.82$\\[1mm]
$\lambda$ & $1$ & $1$\\[1mm]
\hline
\multicolumn{3}{|c|}{$\,$}\\[-4mm] 
\multicolumn{3}{|c|}{Derived Parameters} \\
 \hline
&&\\[-4mm] 
$m_\phi$~[keV] & $37.1$ & $34710$ \\[1mm]
$(\Delta V)^{1/4}$~[keV] & $13.2$ & $12438$\\[1mm]
\hline
\multicolumn{3}{|c|}{$\,$}\\[-4mm] 
\multicolumn{3}{|c|}{Phase Transition} \\
 \hline
&&\\[-4mm] 
$t_*$~[s] & $824$ & $0.0018$ \\[1mm]
$T_*$~[MeV] & $0.040$ & $20$\\[1mm]
$T_{\text{DS},*}$~[MeV] & $0.015$ & $13.8$\\[1mm]
$\alpha$ & $0.01$ & $0.04$\\[1mm]
\hline
\multicolumn{3}{|c|}{$\,$}\\[-4mm] 
\multicolumn{3}{|c|}{Dark Matter} \\
 \hline
&&\\[-4mm] 
$\langle\sigma_{\chi\chi\rightarrow \xi\xi}\, v \rangle$~[$\text{cm}^3s^{-1}$] & $\quad 1.0\times 10^{-26}\quad$ & $\quad 2.5\times 10^{-26}\quad$ \\[1mm]
$T_{\text{nr}}$~[keV] & $20.8$ & $594$\\[1mm]
$T_{\text{FO}}$~[keV] & $2.6$ & $61.7$\\[1mm]
$\Omega_\chi h^2$ & $0.120$ & $0.120$\\[1mm]
$\Delta N_{\text{eff}}$ & $0.10$ & $0.30$\\[1mm]
\hline
\end{tabular}
\end{center}
\vspace{-0.4cm}
\caption{Parameter examples for a Dark Big Bang inducing dark WIMP dark matter. Input and derived Parameters are as defined in Tab.~\eqref{tab:cannibalbp}, and $y'$ is the dark matter/dark radiation coupling of Eq.~\eqref{eq:LDR}.}
\label{tab:rhodarkwimp}
\end{table}

In Fig.~\ref{fig:rhodarkwimp} we depict the temperature-evolution of $\rho_\chi$ for the two parameter examples given in Tab.~\ref{tab:rhodarkwimp} (see Eq.~\eqref{eq:LDS} and Eq.~\eqref{eq:LDR} for the definition of the model Lagrangian). In both cases, the dark WIMPs are produced relativistically and their energy density redshifts as $\rho_\chi\propto T_{\text{DS}}^4\propto T^4$ immediately after the Dark Big Bang. Once the dark WIMPs enter the non-relativistic regime $\rho_\chi$ decreases exponentially due to the Boltzmann suppression factor $\rho_\chi\propto e^{-m_\chi/T_{\text{DS}}}$. This scaling continues as long as $\chi$ remains in thermal equilibrium with the dark radiation plasma. Later -- roughly when the dark WIMP annihilation rate matches the Hubble rate of expansion, $n_\chi\langle \sigma_{\chi\chi\rightarrow \xi\xi}\, v \rangle \sim H$ -- the dark WIMPs freeze out and $Y_\chi$ quickly approaches a final constant value $Y_{\chi,\infty}$. The energy density scales as $\rho_\chi\propto T_{\text{DS}}^3\propto T^3$ after freeze-out. For comparison, we also show the dark matter energy density of $\Lambda$CDM in Fig.~\ref{fig:rhodarkwimp}. As can be seen both dark WIMP scenarios successfully reproduce the correct dark matter density of $\Lambda$CDM in the late Universe.

The thermal freeze-out of dark WIMPs bears resemblance to the case of ordinary WIMPs. In both cases the final relic density scales inversely with the annihilation cross section. However, since the dark WIMPs reside in a colder dark sector, there occurs an additional dependence on the temperature ratio $T_{\text{DS}}/T \sim \alpha^{1/4}$. We find the following approximate scaling relation, $\Omega_\chi \propto \alpha^{1/4}/\langle \sigma_{\chi\chi\rightarrow \xi\xi}\, v \rangle$, i.e.\ there occurs an extra factor $\alpha^{1/4}$ compared to the case of ordinary WIMPs.
Since $\alpha < 1$ (cf.\ Eq.~\ref{eq:alphamax}) this extra factor reduces the annihilation cross section required to reproduce the observed dark matter density for dark WIMPs compared to ordinary WIMPs. Since $\alpha$ cannot be too suppressed due to the lower limit in Eq.~\eqref{eq:alphamin1}, the required dark WIMP annihilation cross section, however, still falls into the same ball park as for ordinary WIMPs. This can directly be verified for the two benchmark points in Tab.~\ref{tab:rhodarkwimp} which both reproduce the observed dark matter density with an annihilation cross section $\langle \sigma_{\chi\chi\rightarrow \xi\xi}\, v \rangle$ in the pb-range.

As discussed in Sec.~\ref{sec:strength} the dark sector energy density can affect BBN and the CMB through its impact on the Hubble expansion rate which is conveniently expressed in terms of an extra contribution $\Delta N_{\text{eff}}$ to the effective neutrino number (as defined in Eq.~\eqref{eq:deltaneffth}).
In addition to the dark WIMPs, the dark sector contains the massless (or very light) dark radiation degree of freedom $\xi$. We can approximate the excess energy density (= the extra energy density as compared to $\Lambda$CDM) after the Dark Big Bang as
\begin{equation}\label{eq:excessenergy}
\rho_{\text{DS}}-\rho_{\text{DM},0}\,a^3 \simeq \frac{\pi^2}{30} g_{\text{DS}}(T_{\text{DS,*}})\, T_{\text{DS,*}}^4\,.
\end{equation}
In the regime, where the dark WIMPs are relativistic, $\chi$ and $\xi$ both contribute to the excess energy density and we can neglect $\rho_{\text{DM},0}\,a^3$ in the above expression (in the relativistic regime $\rho_{\chi}\gg \rho_{\text{DM},0}\,a^3$). In the non-relativistic regime $\rho_{\chi}= \rho_{\text{DM},0}\,a^3$ such that the excess energy density matches the dark radiation density. We can, hence, express the excess energy density in both regimes by the energy density in the relativistic degrees of freedom (which justifies Eq.~\eqref{eq:excessenergy}). By plugging Eq.~\eqref{eq:excessenergy} into Eq.~\eqref{eq:deltaneffth} we obtain the (temperature-dependent) effective number of extra neutrino species~\cite{Nakai:2020oit,Freese:2022qrl},
\begin{equation}\label{eq:neffdw}
\Delta N_{\text{eff}} = 0.63\times \left(\frac{\alpha}{0.1} \right)\left(\frac{10}{g_{\text{eff}}(T_*)} \right)^{1/3}\left(\frac{g_{\text{DS}}(T_{\text{DS,*}})}{g_{\text{DS}}(T_{\text{DS}})} \right)^{1/3}\,,
\end{equation}
where we used Eq.~\eqref{eq:TdT} to express $T_{\text{DS}}/T$ in terms of $\alpha$. As can be seen, $\Delta N_{\text{eff}}$ remains constant or slightly increases (through the mild dependence on $g_{\text{DS}}(T_{\text{DS}})$) as the Universe cools down after the Dark Big Bang. Therefore, the CMB yields the most sensitive probe of the extra energy density predicted in the dark WIMP scenario. The present CMB limit $\Delta N_{\text{eff}}<0.5$ (cf. Eq.~\eqref{eq:deltaneff}) imposes a maximal strength of the Dark Big Bang $\alpha \lesssim 0.1$ (see Sec.~\ref{sec:strength}). 

Eq.~\eqref{eq:neffdw} highlights another important distinction between dark WIMPs and ordinary WIMPs. Dark WIMPs can always be made compatible with $\Delta N_{\text{eff}}$ constraints: if the dark sector is sufficiently cold (i.e.\ $\alpha$ sufficiently small) they only contribute a small fraction of a degree of freedom to $\Delta N_{\text{eff}}$. Ordinary WIMPs, on the other hand, share the temperature of the SM radiation bath. They contribute $\Delta N_{\text{eff}}\gtrsim 1$ to the effective neutrino number as long as they are relativistic. Therefore, contrary to dark WIMPs, ordinary WIMPs are tightly constrained by BBN. In particular, because the observed $^4$He fraction is inconsistent with a full additional neutrino species at BBN (see e.g.~\cite{Cyburt:2015mya,Fields:2019pfx}), ordinary WIMPs need to be non-relativistic at the time of BBN. In contrast, dark WIMPs can evade BBN bounds if $\alpha$ is sufficiently small. They only need to become cold before washing out structures by their free-streaming (see Sec.~\ref{sec:wdm}), i.e.\ at temperatures (dark sector temperatures) of $T>7.8\:\text{keV}$ ($T_d \gtrsim \text{keV}$).\footnote{The dark sector temperature can be up to about one order of magnitude smaller than the visible sector temperature (cf.~Eq.~\eqref{eq:alphamin1}).} As a consequence, dark WIMP dark matter can be realized for significantly lower masses compared to standard WIMP dark matter. Approximately, we have
\begin{equation}
m_\chi \gtrsim  \begin{cases}
\text{MeV}\quad &\text{(WIMPs)}\,,\\
\text{keV}\quad &\text{(dark WIMPs)}\,.
\end{cases}
\end{equation}

Let us now turn to the prospects of revealing the dark WIMP scenario in future observations. In contrast to ordinary WIMPs, dark WIMPs in a decoupled dark sector do not yield any direct or indirect dark matter detection signals. However, the prediction of a fractional contribution to $\Delta N_{\text{eff}}$ renders the dark WIMP scenario very exciting from an observational point of view. For the two dark WIMP benchmark examples of Tab.~\ref{tab:rhodarkwimp} with $\alpha= \mathcal{O}(0.01)$ all present cosmological constraints are satisfied. But the resulting $\Delta N_{\text{eff}}$ falls in the sensitivity window of next-generation CMB experiments like CMB-S4~\cite{Abazajian:2019eic} and Simons Observatory~\cite{SimonsObservatory:2019qwx}. Intriguingly, the same parameter space which is testable through $\Delta N_{\text{eff}}$ predicts a strong gravitational wave signal in reach for future observatories (as we will show in Sec.~\ref{sec:gravitywaves}). Hence, dark WIMPs from a Dark Big Bang offer the thrilling opportunity of a simultaneous, correlated CMB and gravitational wave signal.

\subsection{Heavy Dark-Zilla Dark Matter}\label{sec:darkzilla}

We now turn to the opposite regime $m_\chi\gg (\Delta V)^{1/4}$. In fact, we will mostly focus on ultra-heavy dark matter particles with masses $m_\chi\gg 10^{10}\:\text{GeV}$ which we dub ``dark-zillas'' (since they represent a dark sector version of WIMPzillas~\cite{Kolb:1998ki})

While it may naively seem that the production of ultra-heavy particles in the Dark Big Bang phase transition is kinamatically forbidden, such an argument misses the Lorentz boost of the colliding bubble walls~\cite{Watkins:1991zt}. Since we are considering a decoupled dark sector, the bubbles are not slowed down by pressure when they expand (this scenario goes under the name of `runaway bubbles' in the literature). The Lorentz factor $\gamma_w$ of the colliding bubble walls can be estimated as~\cite{Watkins:1991zt}
\begin{equation}
 \gamma_w\simeq \frac{R_b}{R_0}\simeq 2\,\frac{\Delta V}{m (\Delta\phi)^2}\,\frac{ M_{\text{P}}}{\sqrt{g_*(T_*)}T_*^2 }\,,
\end{equation}
where $R_b$ again denotes the bubble radius at collision which roughly corresponds to the typical distance between bubble nucleation sites $R_b\sim \Gamma^{-1/4}$. In the second step we took the bubble nucleation radius $R_0$ from Eq.~\eqref{eq:nucradius} and used Eq.~\eqref{eq:tstar},~\eqref{eq:Tstar} to express $R_b$ in terms of the temperature of the SM plasma at the Dark Big Bang. The Lorentz factor can take enormous values. For instance for a Dark Big Bang around the time of BBN we find $\gamma_w \sim 10^{20}$.
The Dark Big Bang can (in principle) generate particles with mass up to $m_\chi\sim \gamma_w\, m_\phi$, which opens the intriguing possibility of producing ultra-heavy dark matter. We will see, however, that the efficiency of high-energy particle emission depends on the elasticity of the bubble collisions and on the nature of the final state particles~\cite{Falkowski:2012fb}.

Since -- contrary to the case of light dark matter discussed in Sec.~\ref{sec:lightdm} -- the production of heavy dark matter critically depends on the particle nature, we will consider the complementary cases of real scalar dark matter and of Majorana fermionic dark matter. In both cases we assume that one (or several) dark radiation degree(s) of freedom is (are) also present in the theory which can efficiently be produced in the bubble collisions.\footnote{If such a light degree of freedom would be absent, this would lead to a complicated cosmology since the production of heavy dark matter particles $\chi$ is typically too inefficient to absorb all the energy stored in the bubble walls. As a consequence long-lived remnants of the colliding bubble condensate would emerge from the Dark Big Bang which would evolve in a non-trivial way. In this work, we refrain from considering such a more involved case.} The Lagrangians read
\begin{equation}\label{eq:Ldarkzilla}
\mathcal{L}_{\text{DS}}= \frac{1}{2}\partial_\mu \phi\partial^\mu \phi- V(\phi)+ \mathcal{L}_{\text{DR}} +\begin{cases}
 \frac{1}{2}\partial_\mu \chi\partial^\mu \chi   - y\,\phi^2 \chi^2 - \frac{m_\chi^2}{2}\chi^2 -\kappa \chi^4 &\text{(real scalar),}   \\ 
\frac{i}{2}\bar{\chi}\cancel\partial\chi 
 - y\, \phi \bar{\chi}{\chi}
 - \frac{m_\chi}{2} \bar{\chi}\chi   & \text{(Maj.\ fermion).}
\end{cases}
\end{equation}
where the potential of the tunneling field $V(\phi)$ was defined in Eq.~\eqref{eq:Vphi}, and $\mathcal{L}_{\text{DR}}$ stands for the Lagrangian terms of the dark radiation field(s). Notice that the Lagrangian for the real scalar case agrees with the one considered in the previous sections (cf.~Eq.~\eqref{eq:LDS}). It was restated merely for convenience. The hierarchy $m_\chi \gg m_\phi$ -- which we will assume in this section -- is not stable against radiative corrections (due to diagrams with $\chi$ running in the loop). While this signals a potential fine-tuning issue, the SM already suffers from an analogous hierarchy problem. We shall, therefore, assume that the same (unknown) mechanism which protects the electroweak scale against large radiative corrections stabilizes the hierarchy $m_\chi \gg m_\phi$.

In order to assess the dark matter production by bubble collisions, it is convenient to treat the colliding vacuum bubbles as a classical external field configuration which acts as a source term for the quantum states to which it couples~\cite{Watkins:1991zt}. In~\cite{Falkowski:2012fb}, the evolution of the scalar field configuration during a first-order phase transition has been modelled for two idealized cases describing the colliding bubble condensate,
\begin{itemize}
\item perfectly elastic collisions: the bubble walls are reflected upon collision, before the vacuum pressure makes them approach and collide again -- a process that repeats many times with particle production occurring in each step. This case is realized for $\Delta V \ll V_b$ such that the bubble collisions can temporarily restore a region of the false vacuum phase between the bubble walls.
\item totally inelastic collisions: the bubble walls merge upon collision and efficiently transfer their energy to scalar waves/ particles. This case is realized for $\Delta V \gg V_b$ such that $\phi$ remains in the attraction of the true minimum during the collision.
\end{itemize}
These two scenarios provide useful limiting cases with the true particle production expected to fall between them. The mean number of dark matter particles $N_\chi$ produced per unit area $A$ in the bubble collisions is given as~\cite{Falkowski:2012fb}
\begin{equation}\label{eq:NchiA}
\frac{N_\chi}{A} = \frac{1}{2\pi^2} \int\limits_{\epsilon_{\text{min}}}^{\epsilon_{\text{max}}} d \epsilon f(\epsilon)\int d\Pi_{2} \left| \bar{\mathcal{M}}(\phi\rightarrow \chi\chi)\right|^2\,,
\end{equation}
where $\left| \bar{\mathcal{M}}(\phi\rightarrow \chi\chi)\right|^2$ is the squared amplitude (spin-averaged squared amplitude in case of fermionic $\chi$) for the decay $ \phi\rightarrow \chi\chi$,\footnote{Here we assume that $\chi$ is pair-produced as suggested by dark matter stability.} while $\Pi_2$ stands for the relativistically invariant 2-body phase space element. The first integral runs over the available invariant masses squared $\epsilon$, where $\epsilon_{\text{min}}$ and $\epsilon_{\text{max}}$ are the minimal and maximal $\epsilon$ imposed by kinematics for producing particles $\chi$,
\begin{equation}
\epsilon_{\text{min}} = 4 m_\chi^2\,,\qquad \epsilon_{\text{max}} = \gamma_w^2 m_\phi^2\,.
\end{equation}
Notice that $\sqrt{\epsilon}$ corresponds to the summed energy of the two outgoing particles.

The function $f(\epsilon)$ encodes the efficiency of particle production as a function of energy which depends on the details of the bubble collisions. For the limiting cases of perfectly elastic and totally inelastic collisions the following analytic expressions  have been obtained in~\cite{Falkowski:2012fb},
\begin{equation}\label{eq:fepsilon}
f(\epsilon)= \begin{cases}
\frac{16 (\Delta\phi)^2 }{\epsilon^2}\,\log\left(
\frac{2\epsilon_{\text{max}}-\epsilon+ 2 \sqrt{\epsilon_{\text{max}}^2-\epsilon\,\epsilon_{\text{max}}}}{\epsilon}
\right) & \text{(elastic),}\\
\frac{4 (\Delta\phi)^2 }{\epsilon^2}\,
\left(
\frac{\left(\epsilon- m_\phi^2\right)^2}{m_\phi^4}+\frac{m_\phi^2}{\epsilon_{\text{max}}} 
\right)^{-1}
\,\log\left(
\frac{2\epsilon_{\text{max}}+\epsilon+ 2 \sqrt{\epsilon_{\text{max}}^2+\epsilon\,\epsilon_{\text{max}}}}{\epsilon}
\right)& \text{(inelastic).}
\end{cases}
\end{equation}

In the next step, the matrix elements $\phi\rightarrow  \chi\chi$ for the dark matter pair-production need to be derived. We obtain\footnote{Notice that one must include a factor of $1/2$ in the phase-space integral due to the occurrence of two identical particles in the final state.},
\begin{equation}\label{eq:matrix}
\int d\Pi_{2} \left| \bar{\mathcal{M}}(\phi\rightarrow \chi\chi)\right|^2
=
\begin{cases}
\frac{y^2\,(\Delta\phi)^2}{\pi}\sqrt{1-\frac{4 m_\chi^2}{\epsilon}}   & \text{(real scalar),}\\
\frac{y^2}{2\pi}\epsilon\left(1-\frac{4 m_\chi^2}{\epsilon}\right)^{3/2}   & \text{(Maj.\ fermion).}
\end{cases}
\end{equation}
Notice the additional factor $\epsilon$ in the fermionic case. 
We can now determine $N_\chi/A$ by plugging Eq.~\eqref{eq:fepsilon} and Eq.~\eqref{eq:matrix} into Eq.~\eqref{eq:NchiA}.
In the integral, one can see that the extra factor of $\epsilon$ in the case of fermions in Eq.~\eqref{eq:matrix} is responsible for the fact that heavy fermions are produced much more efficiently than heavy scalars in a first-order phase transition~\cite{Falkowski:2012fb}.  

The number of dark-zillas produced at the walls of the colliding bubbles can be translated to the dark-zilla number density immediately after the Dark Big Bang,
\begin{equation}
n_\chi(T_*) \simeq \frac{N_\chi}{A} \frac{3}{2\,R_b} \sim \frac{N_\chi}{A} \frac{3\,\Gamma^{1/4}}{2} \sim \frac{N_\chi}{A} \frac{\sqrt{g_{\text{eff}}(T_*)}\, T_*^2}{M_P}\,,
\end{equation}
where $R_b$ again stands for the bubble radius at collision. In the last two steps we employed Eq.~\eqref{eq:Rb}, Eq.~\eqref{eq:tstar} and Eq.~\eqref{eq:Tstar} in order to express $R_b$ in terms of $T_*$.

The dark-zilla energy density can be written as,
\begin{equation}
\rho_\chi = n_\chi \langle E_\chi \rangle\,,
\end{equation}
where $\langle E_\chi \rangle$ denotes the mean dark-zilla energy. Immediately after the Dark Big Bang one has,
\begin{equation}\label{eq:Echimean}
\langle E_\chi \rangle_* = \frac{\int\limits_{\epsilon_{\text{min}}}^{\epsilon_{\text{max}}} d \epsilon \sqrt{\epsilon}\,f(\epsilon)\int d\Pi_{2} \left| \bar{\mathcal{M}}(\phi\rightarrow \chi\chi)\right|^2}{\int\limits_{\epsilon_{\text{min}}}^{\epsilon_{\text{max}}} d \epsilon f(\epsilon)\int d\Pi_{2} \left| \bar{\mathcal{M}}(\phi\rightarrow \chi\chi)\right|^2}\,.
\end{equation}
Due to the large Lorentz boost of the colliding bubble walls, the produced dark-zillas can be very energetic.

Validity of the described formalism of particle production in the phase transition requires $\rho_{\chi,*}< \rho_\phi$. If $\rho_{\chi,*}$ derived from the above expressions exceeds $\rho_\phi$ this signals an inconsistency since the total energy density released into particles at the Dark Big Bang can obviously not be larger than $\rho_\phi$ by means of energy conservation. What physically happens if $\rho_{\chi,*}$ approaches $\rho_\phi$ is that particle production in the phase transitions becomes so efficient that it backreacts on the field configuration of the vacuum bubbles~\cite{Falkowski:2012fb}. Since this backreaction is not accounted for in the described formalism (where the bubble walls are treated as an external source), we can no longer trust the calculation in the regime where $\rho_{\chi,*}$ approaches $\rho_\phi$. In the following we will thus impose $\rho_{\chi,*}< \rho_\phi$ such that the formalism can safely be applied.

We will, furthermore, assume that the fraction of vacuum energy which is not transferred to dark-zillas is converted into dark radiation, i.e.\ we set
\begin{equation}
\rho_{\text{DR},*} = \rho_\phi - \rho_{\chi,*}\,.
\end{equation}
This assumption is plausible since light degrees of freedom are efficiently produced by the Dark Big Bang and should thus be the dominant final state. The dark radiation particles $\xi$ will quickly thermalize by self-scattering and form a dark plasma characterized by the dark sector temperature $T_{\text{DS}}$. 

After the Dark Big Bang, the total number of dark-zillas in the Universe is (typically) conserved, and, hence, the number density scales as $n_\chi\propto a^{-3}$. This is because -- for a fixed $\rho_\chi$ -- the dark-zilla number density is inversely proportional to their (ultra-heavy) mass, $n_\chi\propto 1/m_\chi$. Hence, the rate at which dark-zillas find an interaction partner to undergo pair-annihilation is negligibly small. The relic abundance of dark-zillas is thus simply given by,
\begin{equation}
Y_{\chi,\infty} \simeq \frac{n_{\chi,*}}{s_*}
\sim \frac{N_\chi}{A}\frac{2}{\sqrt{g_{\text{eff}}(T_*)}\, M_P\, T_*}\,,
\end{equation}
where $s_*$ denotes the visible sector entropy at the time of the Dark Big Bang. Finally, the dark-zilla relic density is obtained via Eq.~\eqref{eq:Omegachi}.

\begin{figure}[htp]
\begin{center}
\includegraphics[width=0.55\textwidth]{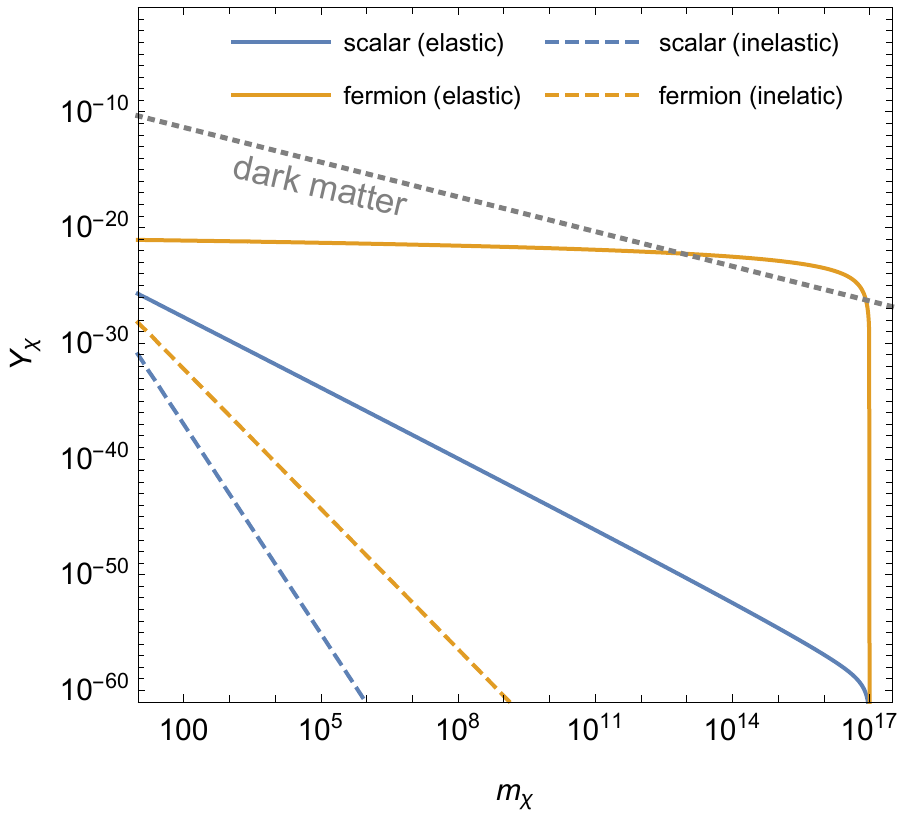}
\end{center}
\caption{Relic abundance of heavy dark-zillas produced by the bubble collisions in a Dark Big Bang first-order phase transition as a function of their mass (input parameters: $T_*=1\:\text{GeV}$, $m_\phi=2\:\text{GeV}$, $\Delta\phi=0.5\:\text{GeV}$, $y=0.01$, $\gamma_w=10^{17}$). The blue (orange) lines refer to real scalar (Majorana fermion) dark-zillas produced by perfectly elastic (solid lines) or totally inelastic (dashed lines) bubble collisions. For comparison, the relic abundance which corresponds to the observed dark matter density is also shown (gray short-dashed line).}
\label{fig:Yzilla}
\end{figure}

In Fig.~\ref{fig:Yzilla} we depict $Y_{\chi,\infty}$ for a typical Dark Big Bang phase transition at the GeV-scale. There occur four subcases in which $\chi$ is either identified with a real scalar or a Majorana fermion and in which we assume perfectly elastic or totally inelastic bubble collisions respectively. Also shown is the required $Y_{\chi,\infty}$ to reproduce the observed dark matter density. It can be seen that the production of heavy $\chi$ is more efficient (i) for fermionic $\chi$ (ii) for elastic bubble collisions. The reason that heavy fermions are produced more efficiently lies in the additional factor of $\epsilon$ in their matrix element compared to the scalar case (cf.~Eq.~\eqref{eq:matrix}). Similarly, the spectral function $f(\epsilon)$ is larger by a factor $\epsilon^2/m_\phi^4$ for elastic bubble collisions compared to inelastic ones (cf.~Eq.~\eqref{eq:fepsilon}). As a consequence, among the four subcases, only Majorana fermion final states
produced by elastic bubble collisions reach a sufficiently large relic abundance to account for all dark matter in the Universe. We will, therefore, focus on this case in the following.

We do note other possibilities however: In this section we only consider heavy real scalar and heavy Majorana fermion final states. Among those, only the Majorana fermions are sufficiently produced by bubble collision to account for the dark matter. We note, however, that Dirac fermions and vector bosons also constitute viable dark-zilla dark matter candidates. We refrain from considering these additional cases in this work since they would give rise to a similar phenomenology as for the Majorana fermionic dark-zillas.

If we identify the dark-zillas with Majorana fermions from elastic collisions, the high-energy Fourier modes of the colliding bubbles contribute substantially to particle formation. This leads to the nearly constant $Y_{\chi,\infty}$ as a function of $m_\chi$ which we observe in Fig.~\ref{fig:Yzilla} (unless $m_\chi$ approaches the maximal available energy $\sqrt{\epsilon_{\text{max}}}=\gamma m_\phi$, where an additional suppression sets in). Since $\Omega_\chi \propto m_\chi\,Y_{\chi,\infty}$, the dark-zilla relic density even grows with $m_\chi$. Fig.~\ref{fig:Yzilla} suggests dark-zillas as heavy as $m_\chi\sim 10^{13}\:\text{GeV}$ can account for all dark matter in the Universe -- a remarkable observation given the phase transition occurred at an energy scale as low as $T_*=1\:\text{GeV}$. In contrast to many other dark matter production mechanisms, a first-order phase transition is thus able to efficiently produce high-mass particles.

As we already noted, the dark-zillas may carry substantial amounts of kinetic energy immediately after the Dark Big Bang. In order to constitute a viable dark matter candidate, they need to lose their kinetic energy quickly enough such that their free-streaming does not spoil structure formation. 

The dark-zillas may lose their kinetic energy through (i) collisions of the dark-zillas with dark radiation particles $\xi$, (ii) redshifting as the Universe expands. The collision rate of dark-zillas can be estimated as $\Gamma_{\text{collision}}\sim \langle\sigma_{\chi\xi\rightarrow \chi\xi}\, v \rangle \,n_\xi \sim T_d^3 /m_\chi^2$ for dimensional grounds, while the redshifting is controlled by the Hubble rate $H\sim T^2 / M_P$. Because dark-zillas need to be ultra-heavy to account for the dark matter of the Universe (see above), we typically find $H \gg \Gamma_{\text{collision}}$. For instance, for the phase transition from Fig.~\ref{fig:Yzilla}, we find $\Gamma_{\text{collision}}/H\lesssim 10^{-8}$ if we impose the correct dark matter relic density. Assuming, therefore, that redshifting is the dominant process, we can approximate the temperature of the Universe $T_{\text{nr}}$ at which the dark-zillas become non-relativistic,
\begin{equation}
 T_{\text{nr}} \simeq \left(\frac{g_{\text{eff}}(T_*)}{g_{\text{eff}}(T)}\right)^{1/3}\frac{a_*}{a_{\text{nr}}}\,T_* \simeq \left(\frac{g_{\text{eff}}(T_*)}{g_{\text{eff}}(T)}\right)^{1/3}\frac{m_\chi}{\langle E_\chi \rangle_*}\,T_*\,,
\end{equation}
where we employed entropy conservation in the first step and the redshifting of $\langle E_\chi \rangle \propto 1/a$ in the second step (for as long as $\chi$ is relativistic). The mean dark-zilla energy at the Dark Big Bang $\langle E_\chi \rangle_*$ can be obtained from Eq.~\eqref{eq:Echimean}.
As an estimate of the constraint from structure formation, we can apply the WDM bound derived in Eq (\ref{eq:TNR}) 
which requires $T_{\text{nr}}>7.8\:\text{keV}$.

\begin{table}[htp]
\begin{center}
\begin{tabular}{|l|c|}
\hline
\multicolumn{2}{|c|}{$\,$}\\[-4mm] 
\multicolumn{2}{|c|}{Input Parameters} \\
 \hline
&\\[-4mm] 
$m_\chi$~[GeV] & $2.4\times 10^{11}$\\[1mm]
$m$~[GeV] & $0.76$ \\[1mm]
$\mu$~[GeV] & $1.19$ \\[1mm]
$y$ & $0.02$ \\[1mm]
$\lambda$ & $1$ \\[1mm]
\hline
\multicolumn{2}{|c|}{$\,$}\\[-4mm] 
\multicolumn{2}{|c|}{Derived Parameters} \\
 \hline
&\\[-4mm] 
$m_\phi$~[GeV] & $1.14$  \\[1mm]
$(\Delta V)^{1/4}$~[GeV] & $0.41$ \\[1mm]
\hline
\multicolumn{2}{|c|}{$\,$}\\[-4mm] 
\multicolumn{2}{|c|}{Phase Transition} \\
 \hline
&\\[-4mm] 
$t_*$~[$\mu$s] & $1.4$  \\[1mm]
$T_*$~[GeV] & $0.46$ \\[1mm]
$T_{\text{DS},*}$~[GeV] & $0.54$ \\[1mm]
$\alpha$ & $0.03$ \\[1mm]
\hline
\multicolumn{2}{|c|}{$\,$}\\[-4mm] 
\multicolumn{2}{|c|}{Dark Matter} \\
 \hline
&\\[-4mm] 
$\rho_{\chi,*}/\rho_{\text{DR},*}$ & $0.001$  \\[1mm]
$\langle E_\chi \rangle_*/m_\chi$ & $1.9\times 10^4$\\[1mm]
$T_{\text{nr}}$~[keV] & $61$ \\[1mm]
$\Omega_\chi h^2$ & $0.120$ \\[1mm]
$\Delta N_{\text{eff}}$ & $0.10$ \\[1mm]
\hline
\end{tabular}
\end{center}
\vspace{-0.4cm}
\caption{Parameter example for dark-zilla dark matter induced by a Dark Big Bang.  The dark-zilla is taken to be a Majorana fermion. The vacuum bubbles nucleated in the phase transition are assumed to collide perfectly elastically. Input and derived parameters are as defined in Tab.~\eqref{tab:cannibalbp}, and the coupling $y$ between the tunneling and dark matter fields is defined in Eq.~\eqref{eq:Ldarkzilla}. }
\label{tab:darkzilla}
\end{table}

In Tab.~\ref{tab:darkzilla} we present a parameter example of dark-zilla dark matter produced in a Dark Big Bang around the GeV-scale. The dark-zillas are chosen to be Majorana fermions (see Eq.~\eqref{eq:Ldarkzilla} for the model Lagrangian) and exhibit a large mass of $m_\chi=2.4\times 10^{11}\:\text{GeV}$. While only a subdominant fraction $\rho_{\chi,*}\sim 10^{-3}\times \rho_\phi$ of the vacuum energy released in the Dark Big Bang is transferred to dark-zillas (the remaining $99.9\%$ going into dark radiation) the dark-zillas can account for all dark matter in the Universe. This result holds under the caveat that the dark-zilla production in the phase transition is reasonably approximated by treating the bubble collisions as elastic.\footnote{Strictly elastic bubble collisions occur in the thin-wall regime of vacuum tunneling. The benchmark point resides at the border of the thin-wall regime such that an inelastic component is expected in the bubble collisions. Precise predictions of $\Omega_\chi$ would, therefore, require dedicated simulations in order to trace the field configuration of the colliding bubbles which source the dark-zillas. Since such simulations go beyond the scope of this paper, we only state the result for the idealized totally elastic case in this work.} The dark-zillas fulfil the structure formation constraint since they become non-relativistic at a temperature $T_{\text{nr}}=61\:\text{keV}$.

While we selected a benchmark point in which the dark-zillas play the role of cold dark matter, we note that the temperature of the non-relativistic transition varies vastly within the dark-zilla parameter space. Especially, there also exist parameter combinations in which the heavy dark-zillas only become non-relativistic at keV-temperatures and, therefore, act as WDM. Their phase-space distribution would be rather unique and very distinct from other WDM candidates (like sterile neutrinos or thermal relics) since their energy spectrum is determined by the bubble dynamics during the phase transition. While the matter power spectrum of warm dark-zillas would exhibit a small-scale cutoff similar to thermal WDM, the shape of the cutoff is (mildly) sensitive to the dark matter phase-space distribution (see e.g.~\cite{Ballesteros:2020adh,Dienes:2021cxp}). It would be interesting to explore the corresponding impact on the small-scale structure through hydrodynamic simulations. Intriguingly, if future small-scale structure observations could test the phase-space distribution of warm dark-zillas, this could yield a direct experimental probe of their origin in a Dark Big Bang phase transition.

In addition to any potential WDM signal, the dark-zilla scenario induces a non-vanishing $\Delta N_{\text{eff}}$. The main contribution to $\Delta N_{\text{eff}}$, however, does not come from the dark-zillas themselves, but rather from the dark radiation degree(s) of freedom which are dominantly produced by the Dark Big Bang. Since the dark radiation (typically) thermalizes, we can employ Eq.~\eqref{eq:neffdw} to determine $\Delta N_{\text{eff}}$. For the parameter example of Tab.~\ref{tab:darkzilla} we obtain $\Delta N_{\text{eff}} = 0.10$ -- a value which is in reach for future CMB missions.

In contrast to the dark radiation, the dark-zillas themselves will be more difficult to detect experimentally. An interesting possibility, however, occurs if the symmetry which stabilizes the dark-zillas is very weakly broken (for instance through quantum gravity effects). In this case the dark-zillas could decay into SM states with an extremely large lifetime (larger than the age of the Universe). Even such a highly suppressed decay could give rise to spectacular dark-zilla signatures in the spectrum of ultra-high energy cosmic rays, neutrinos or gamma rays (see e.g.~\cite{Ellis:1990nb,Kuzmin:1997jua,Birkel:1998nx,Berezinsky:1997hy,Aloisio:2007bh,Feldstein:2013kka,Esmaili:2013gha,Dudas:2018npp}).

The dark radiation and/or dark-zilla signatures will be accompanied by the gravitational wave signal generated during the first-order phase transition. As we will show in the next section, substantial dark-zillas parameter space is within reach for future gravitational wave detectors. 

\section{Gravity Waves from the Dark Big Bang}\label{sec:gravitywaves}

First-order phase transitions can generate strong gravitational radiation~\cite{Witten:1984rs,Hogan:1986qda} by the collisions of true vacuum bubbles~\cite{Kosowsky:1992rz,Kosowsky:1992vn} as well as sound waves~\cite{Hindmarsh:2013xza,Hindmarsh:2015qta,Hindmarsh:2017gnf} and magneto-hydrodynamic turbulence in the surrounding plasma induced by the expanding bubbles~\cite{Kosowsky:2001xp,Dolgov:2002ra,Caprini:2009yp}. In the Dark Big Bang scenario -- because the tunneling field $\phi$ is decoupled from ordinary matter and radiation -- the vacuum energy is entirely transferred to the expanding bubbles, while interactions with the surrounding SM plasma play no role. Therefore, only the bubble collisions contribute to the gravitational wave signal from the Dark Big Bang. We refer to reader to~\cite{Schwaller:2015tja,Jaeckel:2016jlh,Addazi:2016fbj,Aoki:2017aws,Bai:2018dxf,Baldes:2018emh,Breitbach:2018ddu,Fairbairn:2019xog,Helmboldt:2019pan,Nakai:2020oit,Addazi:2020zcj,Ratzinger:2020koh,Halverson:2020xpg,Borah:2021ocu,Lewicki:2021xku,Adhikari:2022sbh,Jinno:2022fom} for some previous work on the gravitational radiation from dark phase transitions.

For simple dimensional grounds, the total energy density emitted during the phase transition in the form of gravitational waves can be approximated by Eq.~\eqref{eq:rhogwstar}. The frequency spectrum of the gravitational waves, on the other hand, has to be derived from simulations of the colliding bubble condensate. Such simulations rely on simplified modeling of the vacuum bubbles. In the common envelope approximation~\cite{Kosowsky:1992rz,Kosowsky:1992vn,Huber:2008hg}, the stress-energy is assumed to be located in a thin shell at the bubble wall and to disappear upon collision. Gravitational radiation emerges from the uncollided envelope of the spherical bubbles, while the interaction region is ignored. The envelope approximation is expected to apply to totally elastic bubble collisions in which $\phi$ becomes trapped temporarily in the false vacuum within the bubble collision region~\cite{Jinno:2019bxw}. This occurs in the thin-wall regime of vacuum tunneling, i.e.\ when the energy density difference between the true and the false vacuum is small compared to the barrier separating the two~\cite{Hawking:1982ga,Watkins:1991zt,Falkowski:2012fb}. In the opposite thick-wall regime, the tunneling field does, however, not get trapped and rather undergoes oscillations around the true vacuum within the bubble overlap region. Correspondingly, the shear stress after collision does not vanish -- violating the basic assumptions of the envelope approximation. Simulations of bubble collisions in the thick-wall regime have been performed for instance in~\cite{Cutting:2020nla}. The resulting gravitational wave spectrum is found to deviate from the one predicted in the envelope approximation.
However, differences mostly affect the infrared and ultraviolet tails of the spectrum, while the peak amplitude and frequency turn out to be similar in the two types of simulations. Therefore, in order to perform simple estimates of the Dark Big Bang parameters which generate an observable gravitational wave signal, it is sufficient to employ the envelope approximation. 

The frequency spectrum of the emitted gravitational waves in the envelope approximation normalized to the critical density reads~\cite{Kosowsky:1992rz,Kosowsky:1992vn,Huber:2008hg}
\begin{equation}\label{eq:gravityspectrum}
\Omega_{\text{GW}} h^2(f)  = 2.7\times 10^{-6}\;\left(\frac{10}{g_{\text{eff}}(T_*)}\right)^{1/3}\;\left(\frac{H_*}{\beta}\right)^2\,\left(\frac{\alpha}{1+\alpha}\right)^2\, \frac{3.8\left(f/f_{\text{peak}}\right)^{2.8}}{1+2.8\left(f/f_{\text{peak}}\right)^{3.8}}
\,,
\end{equation}
where the first two terms on the right-hand side account for the redshift from the time of the Dark Big Bang until today. Furthermore, we have assumed that the bubbles propagate at the speed of light and that the surrounding plasma does not inflict any friction on the expanding bubble walls. This is justified due to the decoupling of the dark and the visible sector.

The expected gravitational wave spectrum corresponds to a (smoothly) broken power law with a maximum at the redshifted peak frequency $f_{\text{peak}}$. The latter is determined as,
\begin{equation}\label{eq:peakf}
f_{\text{peak}}= a_* \,f_{\text{peak,}*} = 1.1 \:\text{nHz}\;\left(\frac{f_{\text{peak,}*}}{H_*}\right)\;\left(\frac{g_{\text{eff}}(T_*)}{10}\right)^{1/6}\;\left(\frac{T_*}{10\:\text{MeV}}\right)\,,\quad\;\; f_{\text{peak,}*} \simeq 0.2\beta\,,
\end{equation}
where $f_{\text{peak,}*}$ is the peak frequency at emission (i.e.\ at the Dark Big Bang) which is extracted from simulations~\cite{Huber:2008hg}. The sharp broken power-law form distinguishes the gravitational wave spectrum of first-order phase transitions from the much smoother spectrum expected from astrophysical sources including mergers of supermassive black hole binaries.

Since the Dark Big Bang must occur during radiation domination (see Fig.~\ref{fig:alpharange}), we can set $\alpha\ll 1$. Furthermore, the duration of the phase transition $\beta^{-1}$ is fixed to be 1/8 of a Hubble time (cf.~Eq.~\eqref{eq:beta}). This allows us to simplify the expressions in Eq.~\eqref{eq:gravityspectrum}, Eq.~\eqref{eq:peakf}. We arrive at the following gravitational wave spectrum and peak frequency for a Dark Big Bang phase transition,
\begin{align}\label{eq:gwspectrumDBB}
\Omega_{\text{GW}} h^2(f)  &\simeq 4.2\times 10^{-8}\;\alpha^2\;\left(\frac{10}{g_{\text{eff}}(T_*)}\right)^{1/3}\; \frac{3.8\left(f/f_{\text{peak}}\right)^{2.8}}{1+2.8\left(f/f_{\text{peak}}\right)^{3.8}}\,,\nonumber\\
f_{\text{peak}}&\simeq 1.8 \:\text{nHz}\;\left(\frac{g_{\text{eff}}(T_*)}{10}\right)^{1/6}\;\left(\frac{T_*}{10\:\text{MeV}}\right)
\,.
\end{align}
In this work we are mainly focusing on Dark Big Bangs around or after the time of BBN. As evident from Eq.~\eqref{eq:gwspectrumDBB} the corresponding gravitational wave spectrum is expected to peak around nHz-frequencies. Experimentally, the nHz-band is covered by pulsar timing array (PTA) experiments which aim at detecting gravitational-wave induced variations in the arrival time of the pulses emitted by millisecond pulsars. Currently, among the most sensitive PTA experiments searching for gravitational waves are the North American Nanohertz Observatory for Gravitational Waves (NANOGrav)~\cite{NANOGrav:2020bcs}, the European Pulsar Timing Array (EPTA)~\cite{Chen:2021rqp}, the Parkes Pulsar Timing Array (PPTA)~\cite{Goncharov:2021oub} and the Indian Pulsar Timing Array~\cite{Tarafdar:2022toa}. These join their efforts as the International Pulsar Timing Array (IPTA)~\cite{Antoniadis:2022pcn}. 

Intriguingly, NANOGrav has recently found evidence for a stochastic common-spectrum process in its 12.5-year dataset which affects pulsar timing residuals -- a signal which was later confirmed by PPTA and EPTA. While stochastic gravitational waves provide a plausible explanation for the signal, proof of the characteristic quadrupolar Hellings-Downs correlations~\cite{Hellings:1983fr} is, however, still withstanding. In the near future, sensitivity improvements are expected from the joined IPTA project. Furthermore, the Square Kilometre Array (SKA)~\cite{Dewdney:2009} is currently constructed and will join the search for gravitational waves towards the end of this decade.

\begin{figure}[htp]
\begin{center}
\includegraphics[height=6.91cm]{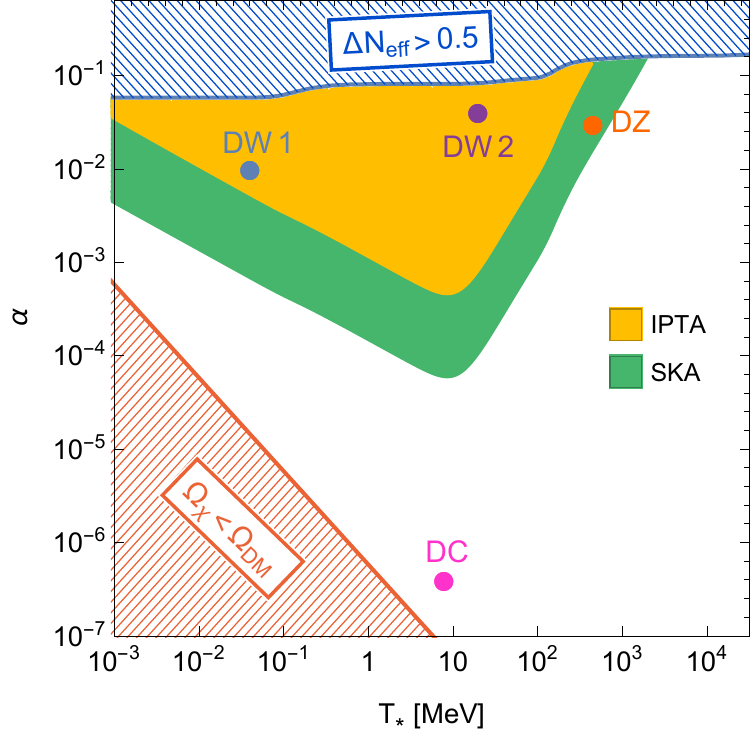}
\hspace{0.02\textwidth}
\includegraphics[height=7cm]{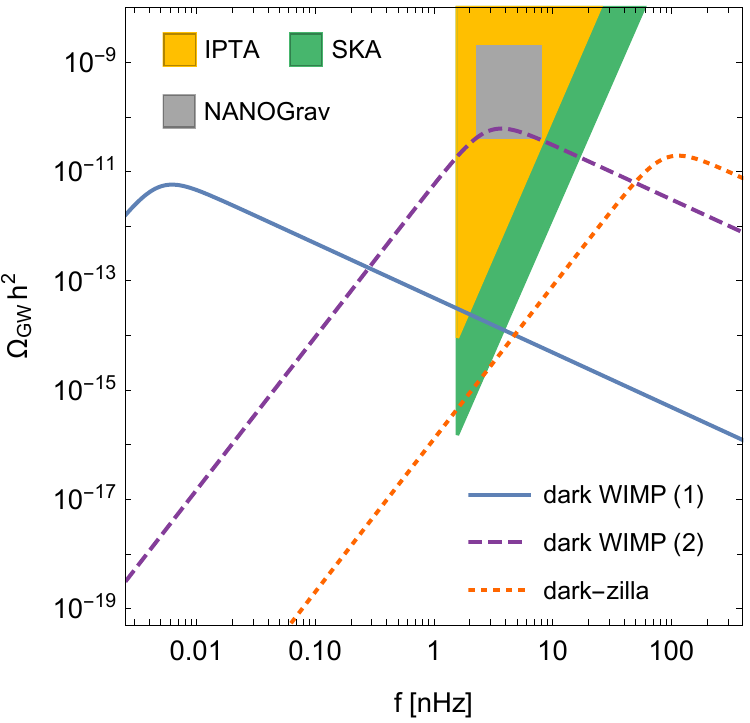}
\end{center}
\caption{Left panel: Projected sensitivity of the IPTA and SKA pulsar timing arrays to a gravitational wave signal from the Dark Big Bang (colored regions). The hatched regions at the bottom and the top of the figure are excluded by a too low dark matter density or by a too large $\Delta N_{\text{eff}}$. The parameter examples of dark cannibal (DC), dark WIMP (DW1 and DW2) and dark-zilla (DZ) dark matter studied in Sec.~\ref{sec:dmproduction} (Tab.~\ref{tab:cannibalbp}-~\ref{tab:darkzilla}) are indicated by the points in the figure. Right panel: Gravitational wave spectrum for the dark WIMP and dark-zilla parameter examples (indicated by the points in the left panel). Also shown are the projected sensitivities of IPTA and SKA as well as the approximate amplitude and frequency of the tentative NANOGrav signal.}
\label{fig:gw}
\end{figure}

The gravitational wave signal from the Dark Big Bang is entirely determined by the strength $\alpha$ and the temperature $T_*$ of the Dark Big Bang (cf.~Eq.\eqref{eq:gwspectrumDBB}). In Fig.~\ref{fig:gw} (left panel) we depict the sensitivity reach of IPTA and SKA in terms of these two parameters.\footnote{The sensitivity reach was derived by requiring that $\Omega_{\text{GW}} h^2(f)$ induced by the Dark Big Bang crosses the experimental sensitivity curves in the $f$-$\Omega_{\text{GW}} h^2$-plane provided in~\cite{Schmitz:2020syl}.} It can be seen that the PTA experiments will cover a substantial part of the parameter space in which the Dark Big Bang can account for all dark matter. The parameter examples of dark WIMP and dark-zilla dark matter induced by a Dark Big Bang around or after the time of BBN, which we studied in Sec.~\ref{sec:dmproduction} -- indicated by the points in the figure -- are fully within reach for the upcoming gravitational wave searches. Since these also exhibit a measurable $\Delta N_{\text{eff}}=0.1-0.3$ (see Tab.~\ref{tab:rhodarkwimp} and Tab.~\ref{tab:darkzilla}) this raises the exciting prospect of a correlated gravitational wave and CMB signal. Dark cannibal scenarios, on the other hand, tend to give a weaker gravitational wave signal since the dark matter is not accompanied by dark radiation in this case (and $\alpha$ is correspondingly lower, cf.~Tab.~\ref{tab:cannibalbp}). We remind the reader, however, that dark cannibals give rise to complementary experimental probes -- for instance related to the shape of dark matter halos -- due to their sizeable self-interaction cross section. 

In the right panel of Fig.~\ref{fig:gw} we depict the gravitational wave spectrum for the dark WIMP and dark-zilla benchmark points together with the IPTA and SKA sensitivities. In this panel we also indicate the frequency and amplitude of the tentative NANOGrav gravitational wave signal (the PPTA and EPTA signals are very similar). The tentative NANOGrav signal would be consistent with the dark WIMP-induced spectrum (purple curve in the figure). Hence -- as we already pointed out in~\cite{Freese:2022qrl} -- the Dark Big Bang provides an attractive explanation for the NANOGrav observation.

We can conclude that PTAs provide a powerful tool to discover the gravitational radiation induced by the Dark Big Bang. In the most optimistic case, first hints of a Dark Big Bang have already been observed in the NANOGrav, PPTA and EPTA data and the Hellings-Downs correlations are soon to be established.

\section{Conclusion}\label{sec:conclusion}

We have introduced an alternative cosmology in which the Hot Big Bang only produces the visible matter and radiation, while the origins of dark matter lie in a Dark Big Bang -- a first-order phase transition in the dark sector. The Dark Big Bang was found to be consistent with all cosmological constraints. In particular, we proved that the dark matter from the Dark Big Bang exhibits the right adiabatic perturbations required for successful structure formation. These fluctuations are imprinted onto the dark matter precisely at the Dark Big Bang due to small differences in the time of the Dark Big Bang in different patches of the Universe which are caused by the perturbations in the (dominant) radiation plasma. 

Dark Matter from the Dark Big Bang can exhibit some peculiar properties: it may be generated rather late in the Universe -- several days after the onset of primordial nucleosynthesis -- without spoiling the light element abundances. Furthermore, its possible mass range is enormous. We have identified successful scenarios of dark WIMP and dark cannibal dark matter with mass as low as $m_\chi \sim 10\:\text{keV}$, where the correct relic density occurs through a thermal freeze-out in the dark plasma. In contrast, the collisions of true vacuum bubbles during the phase transition can produce viable dark-zilla dark matter as heavy as $m_\chi\sim 10^{12}\:\text{GeV}$ -- even for a Dark Big Bang at the MeV-scale. Such huge masses are possible because the expanding bubble walls do not experience any friction and, therefore, typically reach gigantic Lorentz boosts before they collide -- a direct consequence of the decoupling of the visible and the dark sector.

While conventional direct and indirect dark matter searches are doomed to fail in the Dark Big Bang scenario, there nevertheless occur exciting experimental signatures. In particular, the Dark Big Bang naturally allows for realizations of warm and/ or self-interacting dark matter which can be probed through its imprints on dark matter halos and structures in the Universe. While the light dark matter candidates exhibit a (redshifted) thermal spectrum, the phase space distribution of the ultra-heavy dark-zillas is set by the bubble collisions during the phase transition. Intriguingly, if future warm dark matter searches are able to access the dark matter phase space distribution, a direct test of the Dark Big Bang origin of the dark matter could become feasible.

The dark matter from the Dark Big Bang is often accompanied by a dark radiation density which manifests as an extra contribution $\Delta N_{\text{eff}}$ to the effective neutrino number measured in the CMB. For instance, the dark WIMP and dark-zilla benchmark scenarios studied in this work (see Sec.~\ref{sec:dmproduction}) give rise to $\Delta N_{\text{eff}}=0.1-0.3$. This range of $\Delta N_{\text{eff}}$ is fully within reach for upcoming CMB experiments like CMB-S4~\cite{Abazajian:2019eic} and Simons Observatory~\cite{SimonsObservatory:2019qwx}.

The Dark Big Bang phase transition generates strong gravitational radiation. We derived the gravitational wave frequency spectrum of the Dark Big Bang within the standard envelope approximation. Then, we investigated the sensitivity of ongoing and upcoming pulsar timing array experiments to the gravitational wave signal from the Dark Big Bang. We found that already the ongoing IPTA run~\cite{Antoniadis:2022pcn} (which combines several individual PTA experiments) has an exciting discovery potential for Dark Big Bangs which occur around or after BBN. Intriguingly, a tentative gravitational wave signal by the NANOGrav experiment~\cite{NANOGrav:2020bcs} (included in the IPTA network) could already be interpreted as the first sign of the Dark Big Bang~\cite{Freese:2022qrl}. The upcoming SKA~\cite{Dewdney:2009} will further drastically increase the sensitivity of experimental searches for the Dark Big Bang. In addition, there are ongoing efforts to lower the frequency threshold of PTAs~\cite{DeRocco:2022irl} which would allow the measurement of the peak in the gravitational wave spectrum for Dark Big Bang phase transitions which occur long after BBN. Excitingly, the strength of the gravitational wave signal from the phase transition simultaneously determines the Dark Big Bang contribution to $\Delta N_{\text{eff}}$. This raises the fascinating prospect of correlated gravitational wave and CMB signals of the Dark Big Bang which can be probed at the most sensitive near-future laboratories.

In the future, it will be interesting to generalize our results to different variants of Dark Big Bang scenarios in several ways.  First, one can generalize the Dark Big Bang beyond a first-order phase transition. While the gravitational wave production described in this paper requires the bubble collisions of a first-order phase transition, other parts of our analysis -- including aspects of structure formation as well as of the dark matter phenomenology -- would hold for more general Dark-Big-Bang-type scenarios. Furthermore, variations of the discussed scenario including small couplings between the dark and the visible sector could give rise to exciting signatures in cosmic rays (see Sec.~\ref{sec:darkzilla}), dark matter searches, as well as in the CMB and BBN.

\section*{Acknowledgements}
We would like to thank Mike Boylan-Kolchin, Martina Gerbino, Massimiliano Lattanzi, Benjamin Lehmann, Barmak Shams Es Haghi and Hai-Bo Yu for helpful discussions on topics covered in the manuscript. K.F. is Jeff \& Gail Kodosky Endowed Chair in Physics at the University of Texas at Austin, and K.F. and M.W. are grateful for support via this Chair. K.F. and M.W. acknowledge support by the U.S. Department of Energy, Office of Science, Office of High Energy Physics program under Award Number DE-SC-0022021 as well as support from the Swedish Research Council (Contract
No. 638-2013-8993).
\bibliography{dm}
\bibliographystyle{h-physrev}

\end{document}